\documentclass[12pt,preprint]{emulateapj}

\usepackage{epsfig}
\usepackage{epstopdf}
\usepackage{subfigure}

\begin{document}

\title{The Black Hole Mass in Brightest Cluster Galaxy NGC 6086}
\author{Nicholas J. McConnell \footnotemark[1], Chung-Pei Ma \footnotemark[1], James R. Graham \footnotemark[1,2], Karl Gebhardt \footnotemark[3], Tod R. Lauer \footnotemark[4], Shelley A. Wright \footnotemark[1], and Douglas O. Richstone \footnotemark[5]}

\footnotetext[1]{Department of Astronomy, University of California at Berkeley, Berkeley, CA; nmcc@berkeley.edu, cpma@berkeley.edu, jrg@berkeley.edu, sawright@berkeley.edu}
\footnotetext[2]{Dunlap Institute for Astronomy \& Astrophysics, University of Toronto, Toronto, Ontario}
\footnotetext[3]{Department of Astronomy, University of Texas at Austin, Austin, TX; gebhardt@astro.as.utexas.edu}
\footnotetext[4]{National Optical Astronomy Observatory, Tucson, AZ; lauer@noao.edu}
\footnotetext[5]{Department of Astronomy, University of Michigan at Ann Arbor, Ann Arbor, MI; dor@astro.lsa.umich.edu}

\begin{abstract}
We present the first direct measurement of the central black hole mass, $M_\bullet$, in NGC 6086, the Brightest Cluster Galaxy (BCG) in Abell 2162.
Our investigation demonstrates for the first time that stellar dynamical measurements of $M_\bullet$ in BCGs are possible beyond the nearest few galaxy clusters.  We observed NGC 6086 with laser guide star adaptive optics and the integral-field spectrograph (IFS) OSIRIS at the W.M. Keck Observatory, and with the seeing-limited IFS GMOS-N at Gemini Observatory North.  We combined the IFS data sets with existing major-axis kinematics, and used axisymmetric stellar orbit models to determine $M_\bullet$ and the $R$-band stellar mass-to-light ratio, M$_\star$/L$_R$.  We find $M_\bullet = 3.6^{+1.7}_{-1.1} \times 10^9$ M$_\odot$ and M$_\star$/L$_R = 4.6^{+0.3}_{-0.7}$ M$_\odot$/L$_\odot$ (68\% confidence), from models using the most massive dark matter halo allowed within the gravitational potential of the host cluster.  Models fitting only IFS data confirm $M_\bullet \sim 3 \times 10^9$ M$_\odot$ and M$_\star$/L$_R$ $\sim 4$ M$_\odot$/L$_\odot$, with weak dependence on the assumed dark matter halo structure.  When data out to 19 kpc are included, the unrealistic omission of dark matter causes the best-fit black hole mass to decrease dramatically, to $0.6 \times 10^9$ M$_\odot$, and the best-fit stellar mass-to-light ratio to increase to 6.7 M$_\odot$/L$_{\odot,R}$.  The latter value is at further odds with stellar population studies favoring M$_\star$/L$_R$ $\sim 2$ M$_\odot$/L$_\odot$.  Biases from dark matter omission could extend to dynamical models of other galaxies with stellar cores, and revised measurements of $M_\bullet$ could steepen the empirical scaling relationships between black holes and their host galaxies.\\
\end{abstract}

\pagestyle{plain}
\pagenumbering{arabic}

\maketitle

%
\section{Introduction}
\label{sec:intro}

It is increasingly accepted, both observationally and theoretically, that supermassive black holes 
are ubiquitous at the centers of elliptical galaxies \citep{Magorrian}.  The black hole mass, $M_\bullet$, correlates with various host properties, including bulge luminosity, $L$ \citep[e.g.,][]{KR95,MH03}, and stellar velocity dispersion, $\sigma$ \citep[e.g.,][]{Ferr00,Geb00}.  These empirical correlations have been established from approximately 50 galaxies in which $M_\bullet$ has been determined from motions of stars, gas, or masers under the direct gravitational influence of the central black hole.  Although galaxies with $L_V \sim 10^9 - 10^{11}$ L$_{\odot,V}$ are well-represented in this sample \citep[e.g.,][]{HRix, Gultekin}, there are very few measurements of $M_\bullet$ in the most luminous galaxies.

Brightest Cluster Galaxies (BCGs) are the most luminous galaxies in the
present-day universe ($L_V \sim 10^{10.5} - 10^{11.5}$ L$_{\odot,V}$).
Direct measurements of $M_\bullet$ in these galaxies have been lacking
because very few kinematic studies spatially resolve the black hole radius of influence,
$r_{\rm inf} = GM_\bullet / \sigma^2$.
The $M_\bullet - \sigma$ relation predicts typical values of $r_{\rm inf} \sim 30$ pc in BCGs; predictions from the $M_\bullet - L$ relation are a few times larger.  BCGs' low central surface brightnesses exacerbate the
challenge of obtaining high-quality stellar absorption spectra at angular
scales comparable to $r_{\rm inf}$.  To date, stellar dynamical
measurements of $M_\bullet$ in BCGs have been limited to the nearest groups
and clusters: M87 in Virgo \citep[e.g.,][]{Sargent,GT09} and NGC 1399 in
Fornax \citep{Houghton,Geb07}.  In a small number of BCGs, $M_\bullet$ can
be measured from emission line kinematics in a resolved disk of ionized
gas.  \citet{Bonta} have used STIS on the \textit{Hubble Space Telescope}
(\textit{HST}) to examine disks at the centers of three BCGs beyond 50 Mpc,
reporting two measurements of $M_\bullet$ and one upper limit.

BCGs are distinct from other giant elliptical galaxies in several respects.
Two such distinctions are particularly intriguing with regards to the
evolutionary connections between galaxies and their central black holes.
First, BCGs are preferentially found near the gravitational centers of
galaxy clusters, where cosmological dark matter filaments intersect.
Second, BCG luminosities vary more steeply with $\sigma$ than the canonical
$L \propto \sigma^4$ relationship for elliptical galaxies \citep{OH91};
\citet{Lauer07} have found $L \propto \sigma^7$ for BCGs and other core-profile
galaxies.  The steep relationship between $L$ and $\sigma$ in very massive
galaxies requires one or both of the $M_\bullet - \sigma$ and $M_\bullet -
L$ relationships to differ from the correlations observed in lower-mass
galaxies.  Direct measurements of $M_\bullet$ in a statistically
significant sample of BCGs will reveal the forms of these relationships for
the most massive galaxies, and will help discriminate different
evolutionary scenarios for BCGs.  For instance, \citet{BKMaQ06} have
demonstrated that gas-poor galaxies merging on radial orbits could produce
the steep relation between $L$ and $\sigma$.  With little gas available for
star formation or black hole accretion, the remnant galaxy and black hole
would remain on the same $M_\bullet - L$ relation as the progenitors.
These radial mergers could occur at the intersection of cosmological
filaments.
In one counterexample, \citet{RS09} performed a zoom-in resimulation of a single $10^{15} M_\odot$
galaxy cluster selected from a cosmological $N$-body simulation, and produced a BCG that remained on nearly the same $L - \sigma$ relation as the fainter galaxies.  
A larger sample of resimulated clusters would help assess the relative frequency of radial orbits and
their impact on the scaling relations of BCGs.  
Alternative scenarios for BCG growth, such as early-time major mergers \citep[e.g.,][]{Merritt85,Tremaine90} or ``cannibalism'' of smaller
galaxies \citep[e.g.,][]{OT75,OH77}, potentially could produce lower values of
$M_\bullet$, matching predictions from the $M_\bullet - \sigma$ relationship.  In these scenarios, the final black hole mass could depend upon a number of factors, such as the orbits, gas fractions and disk-to-bulge ratios of merging galaxies.  

In addition to providing clues toward BCG evolution, empirically
establishing the high-mass forms of the $M_\bullet - \sigma$ and $M_\bullet
- L$ relationships will provide new constraints for the number density of
the Universe's most massive black holes.  The most luminous high-redshift
quasars are inferred to host black holes exceeding $10^{10}$ M$_\odot$
\citep[e.g.,][]{Bechtold,Netzer,Vestergaard}, but thus far no such objects
have been detected in the local Universe.  BCGs in nearby Abell clusters
potentially could host black holes with $M_\bullet > 10^{9.5}$
M$_\odot$ \citep{Lauer07}. 

Another motivation for measuring $M_\bullet$ in BCGs is that the faint
centers of these galaxies likely arise from ``core-scouring,'' whereby stars are
ejected from the galatic centers by an in-spiraling pair of supermassive
black holes after a major merger \citep[e.g.,][]{EMO91}.  Given theoretical
expectations for the efficiency of core scouring, a galaxy's past merger
history can be estimated by comparing $M_\bullet$ to the total luminosity
deficit in the core \citep{Lauer07,KB09}. 

In this paper, we report measurements of $M_\bullet$ and the $R$-band
stellar mass-to-light ratio, M$_\star$/L$_R$, in NGC 6086, the BCG of Abell
cluster 2162.  Our investigation marks the first attempt to measure
$M_\bullet$ using stellar dynamics in a BCG beyond Virgo.  Future papers
will describe measurements of $M_\bullet$ in a larger sample of BCGs.  For
BCGs at $\sim 100$ Mpc,  8- to 10-meter telescopes with adaptive
optics (AO) are required to obtain good spectra on $\sim 0.1''$ spatial scales.
Laser guide star adaptive optics (LGS-AO) enables the study of targets
without a bright nearby guide star.  We use integral-field
spectrographs (IFS) to obtain full 2-dimensional spatial coverage, which places
tighter constraints on stellar orbits.  Our orbit models include a dark
matter component in the gravitational potential, as described in
\citet{GT09}.  In this paper, we emphasize methods for pairing IFS data
with axisymmetric orbit models, and for assessing errors in $M_\bullet$ and
M$_\star$/L$_R$.

NGC 6086 is a cD galaxy at the center of Abell 2162.  Like many BCGs, it
exhibits radio emission \citep{LO95}, likely from low-level accretion onto
the central black hole.  
We have derived an effective stellar velocity dispersion of 318 km s$^{-1}$ in NGC 6086, using measurements from Carter, Bridges \& Hau (1999; hereafter \citet{Carter}).  This would correspond to a black hole mass of $9 \times 10^8$ M$_\odot$, if NGC 6086 were to follow the mean $M_\bullet - \sigma$ relation of \citet{Gultekin}.  The $V$-band luminosity of NGC 6086 is $1.4 \times 10^{11}$ L$_{\odot,V}$, from $M_V = -23.11$ in \citet{Lauer07}; the corresponding black hole mass predicted from the mean $M_\bullet - L$ relation of \citet{Gultekin} is $1.3 \times 10^9$ M$_\odot$.
Abell 2162 is a relatively small galaxy
cluster at redshift $z = 0.032$, with a richness class of 0 based on 37
members \citep{Abell89}, and a line-of-sight velocity dispersion of
$302^{+132}_{-58}$ km s$^{-1}$ \citep{Zab}.  NGC 6086 is offset from the average
radial velocity of Abell 2162, by 82 km s$^{-1}$ \citep{Laine}.

This paper is organized as follows.  In \S2, we describe our photometric data of NGC 6086, our IFS observations at Keck and Gemini observatories, and the subsequent data reduction procedures.  In \S3, we describe our procedures for extracting 2-dimensional kinematics from IFS data, and compare our resulting measurements in NGC 6086 with other studies.  In \S4 we review the stellar orbit modeling
procedure.  We also report our measurements of $M_\bullet$ and M$_\star$/L$_R$, and describe how these measurements depend on the assumed dark matter halo profile.  We estimate confidence intervals for $M_\bullet$ and M$_\star$/L$_R$, and discuss both tested and un-tested systematic errors.  In \S5 we compare our results to predictions from the $M_\bullet - \sigma$ and $M_\bullet - L$ relationships, and discuss whether the effect of dark matter on stellar orbit models of NGC 6086 can be generalized to reveal biases in measurements of $M_\bullet$ in other galaxies.  Appendix A contains a detailed description of systematic errors from stellar template mismatch and uncertain PSFs.  Appendix B contains our full set of measured line-of-sight velocity distributions (LOSVDs).

Throughout this paper, we assume $H_0 = 70$ km s$^{-1}$, $\Omega_m = 0.27$,
$\Omega_\Lambda = 0.73$, and an angular-diameter distance of 133 Mpc to NGC
6086.  One arc sec corresponds to 0.64 kpc at this distance; for $\sigma = 318$ km s$^{-1}$, $r_{\rm inf} = 0.066'' \times (M_\bullet / 10^9$ M$_\odot$).

%
\section{Observations}
\label{sec:obs}

\subsection{Photometry}
\label{sec:phot}

We use a combination of $R$-band (0.6 $\mu$m) and $I$-band (0.8 $\mu$m) photometry to constrain the stellar mass profile of NGC 6086.  For radii out to $10''$ we adopt the high-resolution surface brightness profile presented in \citet{Laine}, obtained with WFPC2 on the \textit{Hubble Space Telescope}.  This surface brightness profile has been corrected for the WFPC2 point-spread function (PSF) by applying the Lucy-Richardson deconvolution method \citep{Richardson,Lucy}; specific details of the implementation are described in \citet{Laine}.  

At larger radii out to $86''$ we use $R$-band data from Lauer, Postman \& Strauss (private communication), obtained with the 2.1-m telescope at Kitt Peak National Observatory (KPNO).  The KPNO data have a field-of-view (FOV) of $5.2' \times 5.2'$, which enables accurate sky subtraction.  To create a single surface brightness profile, we assessed the individual profiles from WFPC2 and KPNO data at overlapping radii between $5''$ and $10''$.  We measured the average $R-I$ color for these radii and added it to the WFPC2 profile.  The two profiles were then stitched together such that their respective weights varied linearly with radius between $5''$ and $10''$: the WFPC2 data contribute $100\%$ to the combined profile for $r \leq 5''$ and the KPNO data contribute $100\%$ for $r \geq 10''$.  PSF deconvolution was not necessary for the KPNO data, as they contribute to the combined surface brightness profile at radii well beyond the seeing full-width at half-maximum (FWHM).  Our translation of the WFPC2 profile to $R$-band assumes no $R-I$ color gradient; \citet{Lauer05} find a median color gradient, $\frac{\Delta \left(V-I \right)}{\Delta \rm log \left( \it r \right)}$, of -0.03 magnitudes for BCGs and other core profile galaxies.

%
\begin{figure}
  \epsfig{figure=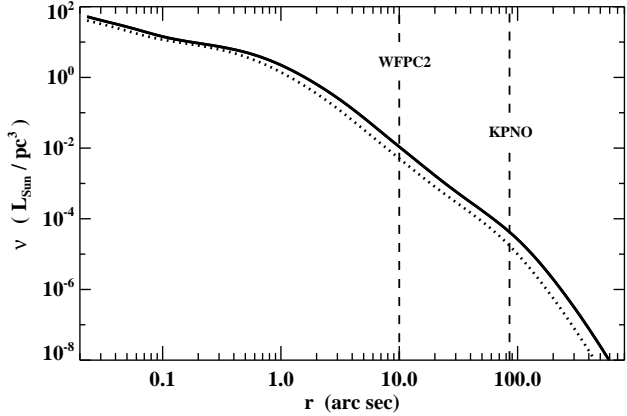,width=3.6in}
 \caption{De-projected $R$-band stellar luminosity density $vs.$ radius along the major axis (solid line) and minor axis (dotted line) of NGC 6086.  The dashed vertical lines mark the outermost extents of photometric data from \textit{HST}/WFPC2 and KPNO.  Luminosity densities beyond $86''$ are derived from a de Vaucouleurs surface brightness profile. The $R$-band surface brightness at $86''$ is 24.3 mag arcsec$^{-2}$ on the major axis.}
\label{fig:lden}
\end{figure}

At radii beyond $1''$, isophotes of NGC 6086 all have major-axis position angles within $5^\circ$ of true north, with an average apparent axis ratio of 0.7.  We adopt $0^\circ$ east of north as the major-axis position angle of NGC 6086, and we assume edge-on inclination.  We deprojected the surface brightness using the procedure of \citet{Geb96}, which assumes spheroidal isodensity contours.  The resulting major- and minor-axis luminosity density profiles are shown in Figure~\ref{fig:lden}.  

\subsection{Spectroscopy}
\label{sec:spec}

We performed integral-field spectroscopic observations of NGC 6086 with OSIRIS \citep{Larkin} on the 10-m W. M. Keck II telescope and GMOS-North (Allington-Smith et al. 2002; Hook et al. 2004) on the 8-m Gemini Observatory North telescope.  The instrument properties and our observations are summarized in Table~\ref{tab:spec}.  Our observations with OSIRIS used the W. M. Keck Observatory laser guide star adaptive optics (LGS-AO) system (Wizinowich et al. 2006; van Dam et al. 2006); the inner component of the resulting $H$-band (1.6 $\mu$m) PSF has an FWHM value of $\approx 0.1''$.  The GMOS data were collected under excellent seeing conditions; images of point sources from the Gemini North Acquisition Camera\footnote{http://www.gemini.edu/sciops/telescopes-and-sites/acquisition-hardware-and-techniques/acquisition-cameras} indicate an $I$-band FWHM of $0.4''$. 

\begin{figure}
  \label{fig:oflux}
  \epsfig{figure=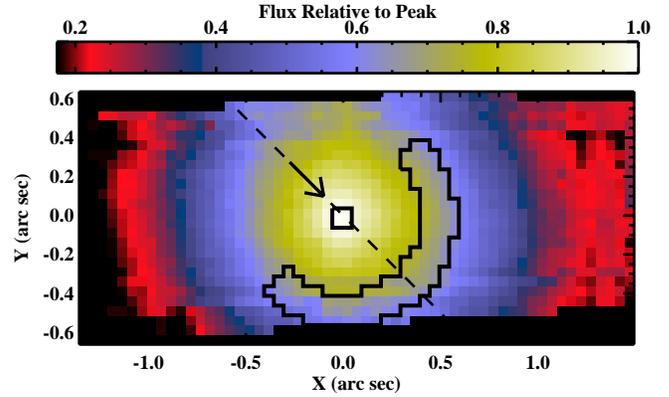,width=3.5in}
\caption{Total $H$-band flux for NGC 6086, using collapsed spectra from OSIRIS.  The dashed line traces the major axis of the galaxy, with the arrow pointing north.  Thick black lines enclose the spatial bins corresponding to the spectra displayed in Figure~\ref{fig:Osample}.}
\label{fig:flux2d}
\end{figure}

In Figure~\ref{fig:flux2d} we display the reduced mosaic of NGC 6086 from OSIRIS, summed over all spectral channels.  Usable data from OSIRIS and GMOS extend to radii of $0.84''$ and $4.9''$, respectively.  For radii out to $30''$ we use major-axis kinematics from 
\citet{Carter},
obtained with the Intermediate Dispersion Spectrograph (IDS)\footnote{http://www.ing.iac.es/Astronomy/instruments/ids/} on the 2.5-m Isaac Newton Telescope.

%
\begin{table*}[htbp]
\begin{center}
\caption{Summary of spectroscopic observations}
\label{tab:spec}
\begin{tabular}[b]{llllllll}  
\hline
Instrument & UT Date & $\lambda$ Range & $\Delta\lambda$  & $\Delta x$ & $t_{int}$ & PA & FWHM\\
&  & (nm) & (nm) & (arc sec) & (s) & ($^\circ$) & (arc sec)\\
\\
(1) & (2) & (3) & (4) & (5) & (6) & (7) & (8)\\
\hline 
\\
OSIRIS & May 13-14, 2008 & $1473 - 1803$ & 0.2 & 0.05 & $9 \times 900$ & $-45$ & 0.10 \\
\\
GMOS & April 25, 2003 & $744 - 948$ & 0.1377 & 0.2 & $5 \times 1200$ & 0 & 0.4 \\
\\
IDS & June 10-16, 1996 & $493 - 573$ & 0.2 & $0.4 \times 3.0$ & 22,800 & 0 & 1.3\\
\\
\hline
\end{tabular}
\end{center}
\begin{small}
\textbf{Notes:}  Column 1: instrument.  OSIRIS (OH-Suppressing Infra-Red Imaging Spectrograph) was used on Keck II with LGS-AO.  GMOS (Gemini Multi-Object Spectrograph) was used on Gemini North.  IDS (Intermediate Dispersion Spectrograph) was used on the Isaac Newton Telescope; here we summarize the observations published by \citet{Carter}.  Column 2: date(s) of observations.  Column 3: observed wavelength range.  Column 4: spectral pixel scale in 3-d data cubes, for OSIRIS and GMOS data.  FWHM spectral resolution, for IDS data.   Column 5: angular spacing of lenslets, for OSIRIS and GMOS data.  Pixel scale along slit $\times$ slit width, for IDS data.  Column 6: number of science exposures $\times$ integration time per exposure.  For IDS data, the total integration time of 6.33 hours is reported from \citet{Carter}.  Column 7: position angle of the long axis for OSIRIS and GMOS, or the slit for IDS, in degrees east of north.  Column 8: PSF FWHM at science wavelengths.
\end{small}
\end{table*}

%

\subsubsection{OSIRIS}
\label{sec:OSIRIS}

OSIRIS is a near-infrared (NIR), integral-field spectrograph built for use with the Keck AO system.  It features 2-dimensional spatial sampling at four scales between $0.02''$ and $0.1''$.  We observed NGC 6086 with the $0.05''$ spatial scale, which provided adequate signal-to-noise and placed several pixels within the radius of influence.  To minimize noise in individual spectra, we used the broad $H$-band filter, which covered several $\Delta\nu = 3$, $^{12}$CO bandheads at observed wavelengths from 1.54 $\mu$m to 1.71 $\mu$m (at $z \approx 0.032$).  We chose to detect $H$-band features instead of the more prominent $\nu = \,$0-2 $^{12}$CO bandhead in $K-band$, which suffered from higher thermal background at an observed wavelength of 2.37 $\mu$m.

We recorded 9 science exposures of the galaxy center and 5 sky exposures of a blank field $50''$ away, for total integration times of 2.25 hr and 1.25 hr, respectively.  Our dithers repeated an ``object-sky-object,'' sequence, such that every science frame was immediately preceded or followed by a sky frame.  
We also recorded spectra of 9 spectral template stars, using the same filter and spatial scale as for NGC 6086.  To measure telluric absorption, we recorded spectra of several A0V stars, covering a range of airmasses similar to those for NGC 6086 and template stars.

%
\begin{figure}
\centering
 \epsfig{figure=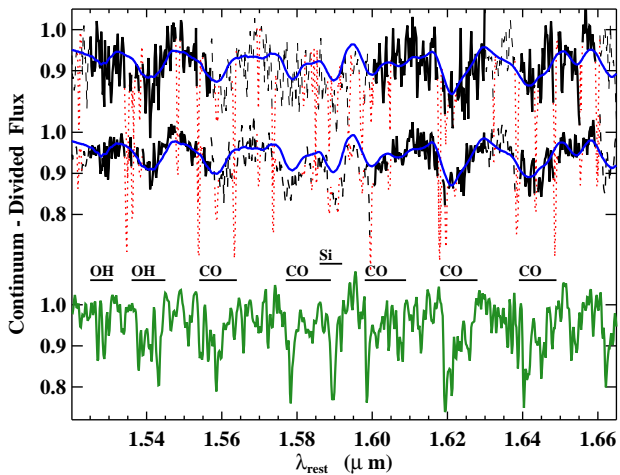,width=3.6in} 
 \caption{\textit{Top:} OSIRIS spectrum of the center of NGC 6086 (0.01 arcsec$^2$; $S/N = 21$).  \textit{Middle:} OSIRIS spectrum $0.49''$ from the center of NGC 6086 (0.28 arcsec$^2$; $S/N = 39$).  The dashed and dotted portions of OSIRIS spectra are excluded from the kinematic fitting, with red dotted portions specifically indicating regions of telluric OH contamination.  Each thick blue line is the M4III template spectrum from OSIRIS, convolved with the best-fit LOSVD for the respective galaxy spectrum.  \textit{Bottom:} Spectrum of template star HD 110964 (M4III), from observations with OSIRIS.  This is the only star used in our final extraction of kinematics from OSIRIS data.}
\label{fig:Osample}
\end{figure}
%

%
\begin{figure}
  \epsfig{figure=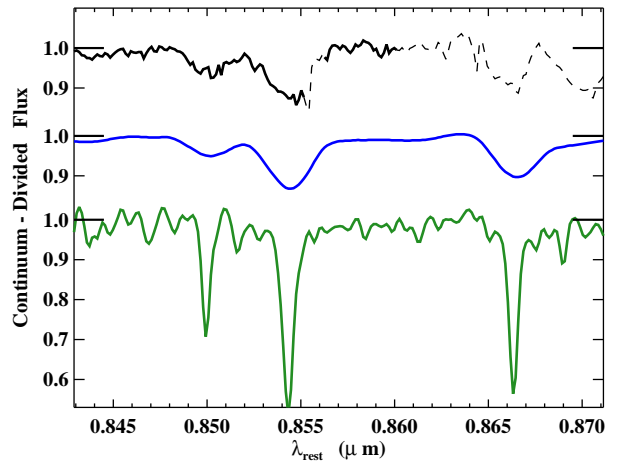,width=3.5in}
 \caption{\textit{Top:} GMOS spectrum of the center of NGC 6086 (0.24 arcsec$^2$; $S/N = 91$).  Dashed portions are excluded from the kinematic fitting.   \textit{Middle:} G9III template spectrum from GMOS, convolved with the best-fit LOSVD.  \textit{Bottom:} Spectrum of template star HD 73710 (G9III), from observations with GMOS.  This is the only star used in our final extraction of kinematics from GMOS data.}
\label{fig:Gsample}
\end{figure}

We used version 2.2 of the OSIRIS data reduction pipeline\footnote{available from the UCLA Infrared Laboratory, at http://irlab.astro.ucla.edu/osiris/pipeline.html}
to subtract sky frames, correct detector artifacts, perform spatial flat-fielding, calibrate wavelengths, generate data cubes with two spatial dimensions ($x$,$y$) and one spectral dimension ($\lambda$), and construct a mosaic of NGC 6086 from multiple data cubes.  The pipeline uses an archived calibration file to perform spectral extraction of the raw spectra across the detector and assemble a data cube; the calibration file was generated by illuminating individual columns of the OSIRIS lenslet array with a white light source.  We used custom routines to remove additional bad pixels from detector images, extract 1-dimensional stellar spectra from 3-dimensional data cubes, and calibrate galaxy and template spectra for telluric absorption.  Although 1-dimensional stellar spectra from OSIRIS comprise an average over many spatial pixels, spatial variations in instrumental resolution are negligible relative to the velocity broadening in NGC 6086: 
$\left(\Delta \sigma_{\rm inst}\right)^2 \sim 5 \times 10^{-3} \, \sigma^2$.

Contamination from telluric OH emission presents a severe challenge for observing faint, extended objects with OSIRIS.  The small field of view ($0.8'' \times 3.2''$ for broadband observations at $0.05''$ per spatial pixel) does not allow for in-field sky subtraction, and subtracting consecutive science and sky frames only provides partial correction, as the relative flux from different vibrational transitions in OH varies on timescales of a few minutes.  After subtracting a sky frame from each science frame, we are forced to discard the spectral channels with strong residual signals from OH, which compose approximately $15\%$ of our spectral range.  
In Figure~\ref{fig:Osample}, we illustrate representative spectra from OSIRIS and distinguish kinematic fitting regions from residual telluric features.  At both ends of the $H$-band spectrum, atmospheric water vapor acts as an additional contaminant.  We have restricted our kinematic analysis to observed wavelengths between 1.48 and 1.73 $\mu$m.

A second challenge for studying the centers of galaxies with OSIRIS is accurate determination of the PSF.  We must construct an average PSF for mosaicked data from several hours of observations over multiple nights, during which seeing conditions and the quality of AO correction can change significantly.  To estimate the PSF, we recorded a one-time sequence of exposures of the LGS-AO tip/tilt star for NGC 6086, using the OSIRIS spectrograph with the same filter and spatial scale settings as for our science frames.  Data cubes were then collapsed along the spectral dimension to produce images of the star.  In Appendix~\ref{app:psf}, we discuss different methods for estimating the PSF, and how PSF uncertainty influences our modeling results.

\subsubsection{GMOS}
\label{sec:GMOS}

GMOS-N is a multi-purpose spectrograph on Gemini North.  GMOS includes an IFS mode, in which hexagonal lenslets divide the focal plane and fibers map the 2-dimensional field to a 1-dimensional slit configuration.  A second set of lenslets samples a field $\sim 60''$ from the science target, allowing for simultaneous sky subtraction.
We observed the center of NGC 6086 in IFS mode with the detector's \textit{CaT} filter, detecting the infra-red CaII triplet at observed wavelengths between 0.87 and 0.90 $\mu$m.  A representative spectrum from the center of NGC 6086 is shown in Figure~\ref{fig:Gsample}.  GMOS data were reduced using version 1.4 of the Gemini IRAF software package\footnote{available from Gemini Observatory, at http://www.gemini.edu/sciops/data-and-results/processing-software}.
This standard pipeline subtracts bias and overscan signals, removes cosmic rays, mosaics data from three CCDs, extracts spectra, corrects throughput variations across fibers and within individual spectra, calibrates wavelengths using arc lamp exposures, computes an average sky spectrum, and performs sky subtraction.  We stored individual spectra from each GMOS exposure, along with their spatial positions relative to the center of NGC 6086, for eventual spatial binning.

With seeing-limited spatial resolution, GMOS poorly resolves the black hole sphere of influence.  Nonetheless, kinematics derived from GMOS provide a good complement to those from OSIRIS.  The CaII triplet region in GMOS spectra has a more clearly-defined continuum than $H$-band spectra, and with less telluric contamination, as is evident from comparing Figures~\ref{fig:Osample} and~\ref{fig:Gsample}.  As a result, line-of-sight velocity distributions (LOSVDs) extracted from GMOS spectra have lower systematic errors than LOSVDs extracted from OSIRIS spectra.  Additionally, the angular region yielding high signal-to-noise spectra from GMOS is four times larger than that for OSIRIS.

%
\section{Kinematics}
\label{sec:kin}

Our dynamical models fit weighted and superposed stellar orbits to LOSVDs extracted from spectroscopic data.  For both OSIRIS and GMOS data, we extract LOSVDs with a Maximum Penalized Likelihood (MPL) technique, which fits an LOSVD-convolved stellar template to each galaxy spectrum.  The LOSVDs are non-parametric, defined at 15 radial velocity bins in our orbit models.  Representative LOSVDs from the central OSIRIS and GMOS bins are shown in Figure~\ref{fig:losvdex}, and the full sets of LOSVDs extracted from OSIRIS and GMOS spectra are presented in Appendix~\ref{app:losvd}.  The MPL fitting method is described in detail in \citet{Geb00b}, \citet{Pinkney}, and \citet{Nowak08}.  Here we describe the specific adjustments made for IFS data of NGC 6086.

%
\begin{figure}
  \epsfig{figure=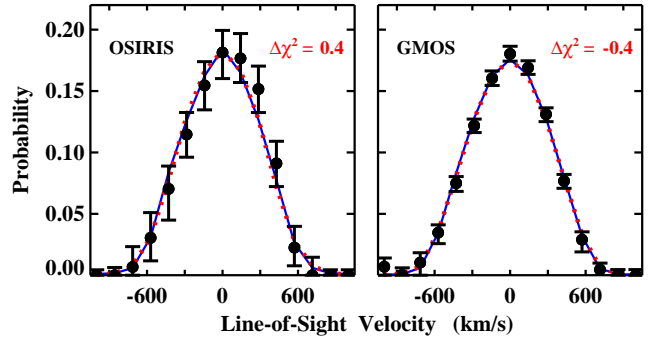,width=3.5in} 
  \caption{Sample LOSVDs for NGC 6086.  \textit{Left:} LOSVD extracted from
    the central spatial region, measured with OSIRIS ($0.1'' \times
    0.1''$).  \textit{Right:} LOSVD extracted from the central spatial region
    measured with GMOS ($\approx 0.55''$ diameter).  The solid blue line in
    each figure is the corresponding LOSVD generated by the
    best-fitting orbit model with the maximum-mass LOG dark matter halo (M$_\star$/L$_R$ =
    4.7 M$_\odot$/L$_\odot$; $M_\bullet = 3.5 \times 10^9$ M$_\odot$; $v_c
    = 500$ kpc; $r_c = 8.0$ kpc).  The dotted red line in each figure is
    from the best-fitting orbit model with no dark matter halo (M$_\star$/L$_R$ =
    6.8 M$_\odot$/L$_\odot$; $M_\bullet = 3.2 \times 10^8$ M$_\odot$).  For each LOSVD, $\Delta\chi^2$ is the difference in the $\chi^2$ statistic for the two models: $\Delta\chi^2 > 0$ indicates that the model including dark matter (solid blue line) is a better fit.  The
    full sets of LOSVDs from OSIRIS, GMOS, and \citet{Carter} are shown in
    Figures~\ref{fig:Olosvd} through~\ref{fig:Ilosvd}.}
\label{fig:losvdex}
\end{figure}

\subsection{Extracting Kinematics from IFS Data}
\label{sec:extract}

In order to attain sufficient signal-to-noise ($S/N$) for effective kinematic extraction, we perform spatial binning on our 2-dimensional grids of spectra from OSIRIS and GMOS.  For the mean-normalized galaxy spectrum $Y$, mean-normalized stellar template $T$, and LOSVD $\mathcal{L}$ from the best fit over $N_c$ spectral channels, we define: 
\begin{equation} S/N  \equiv \left(\sum_{i = 1}^{N_c} \left[ \, Y_i - \left( T *\mathcal{L} \right)_i \, \right]^2 /  \, N_c \right)^{-1/2} 
\label{eq:snr}
\end{equation} 
At the very center of NGC 6086, we spatially bin spectra until $S/N > 20$ is achieved.  This requires binning $2 \times 2$ spatial pixels from OSIRIS; consequently our kinematic data have a central spatial resolution of $0.1''$, similar to the PSF FWHM.  At the center of the GMOS mosaic, we bin 7 hexagonal pixels, corresponding to an approximate diameter of $0.55''$.  The remaining spectra from each dataset are grouped to match angular and radial bins defined within the orbit models, and to maintain $S/N$ between 25 and 40.  Our resulting binning schemes for both OSIRIS and GMOS use only two angular bins on each of the positive (north) and negative (south) sides of the major axis.  The angular bins span $0-36.9^\circ$ and $36.9-90^\circ$ from the major axis.  Axisymmetric models perform LOSVD fitting in one quadrant of the projected galaxy.  Symmetry about the major axis is enforced by co-adding spectra from the positive and negative (east and west) sides of the minor axis, before LOSVD extraction.  LOSVDs extracted from the negative (south) side of the major axis are inverted before being input to the models.  We define systemic velocity relative to the template star separately for OSIRIS and GMOS data.

Additionally, a spectral binning factor is necessary to smooth over channel-to-channel noise in spectra of NGC 6086.  Our final kinematic extraction uses smoothing factors of 30 and 12 spectral pixels for OSIRIS and GMOS spectra, respectively.  These values are chosen by comparing the best-fit LOSVDs from a large range of smoothing factors, and identifying the minimum factor above which LOSVDs in each dataset are stable between $-500$ km s$^{-1}$ and 500 km s$^{-1}$.  Our smoothing values are consistent with the range of optimal values determined by \citet{Nowak08} for near-infrared spectra with $S/N \sim 25-50$. 

$H$-band spectra from OSIRIS contain several absorption features that are potentially useful for kinematic extraction, but some are compromised by incompletely subtracted telluric OH emission lines, which are masked from the fit.  Three broad features are relatively insensitive to the narrow OH lines: the $\nu$ = 3-6 $^{12}$CO bandhead at 1.6189 $\mu$m rest, the $\nu$ = 4-7 $^{12}$CO bandhead at 1.6401 $\mu$m rest, and the $\Delta\nu$ = 2 band of OH between 1.537 and 1.545 $\mu$m rest.  Additionally, the $\Delta\nu$ = 2 OH band between 1.526 and 1.529 $\mu$m rest does not intersect any strong telluric emission features.  We have verified that these four spectral features offer a robust comparison between stellar and galaxy spectra by repeating the fits with a large range of spectral smoothing factors.  When we add other features to the fit, the root-mean-squared residual (RMS, essentially $S/N^{-1}$) becomes unstable to small changes in spectral smoothing.  

To extract LOSVDs from GMOS spectra, we analyzed the $\lambda 8498$ and $\lambda 8542$ lines of CaII.  The third line in the well-known calcium triplet, $\lambda 8662$, is compromised by a 
flat-field artifact and discarded from kinematic analysis.  
The thick black lines in Figure~\ref{fig:Osample} and Figure~\ref{fig:Gsample} indicate the spectral channels used in our final extraction of LOSVDs from OSIRIS and GMOS, respectively.

%
\begin{figure*}[htbp]
  \epsfig{figure=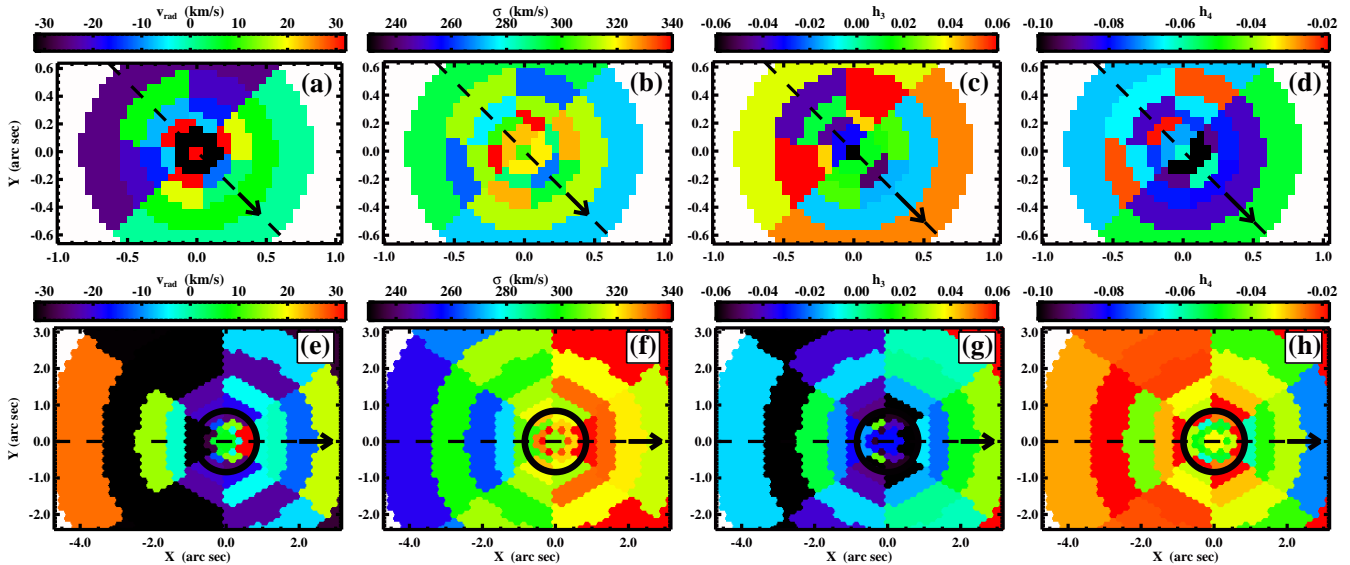,width=7.0in}
 \caption{Two-dimensional kinematics in NGC 6086.  \textbf{(a)} Radial velocity map from OSIRIS.  \textbf{(b)} Velocity dispersion map from OSIRIS.  \textbf{(c)} Map of $h_3$ from OSIRIS.  \textbf{(d)} Map of $h_4$ from OSIRIS.  \textbf{(e)} Radial velocity map from GMOS.   \textbf{(f)} Velocity dispersion map from GMOS.  \textbf{(g)} Map of $h_3$ from GMOS.  \textbf{(h)} Map of $h_4$ from GMOS.  All maps are derived by fitting 4th-order Gauss-Hermite polynomials to non-parametric LOSVDs.  The dashed line in each figure represents the major axis of the galaxy, with the arrow pointing north.  The solid circles in (e) through (h) represent the outermost extent of kinematics from OSIRIS.  The median error values are (a) 41 km s$^{-1}$; (b) 26 km s$^{-1}$; (c) 0.042; (d) 0.024; (e) 20 km s$^{-1}$; (f) 16 km s$^{-1}$; (g) 0.035; and (h) 0.021.}
\label{fig:kinmap}
\end{figure*}
%

%
\begin{figure}[htbp]
 \centering
   \epsfig{figure=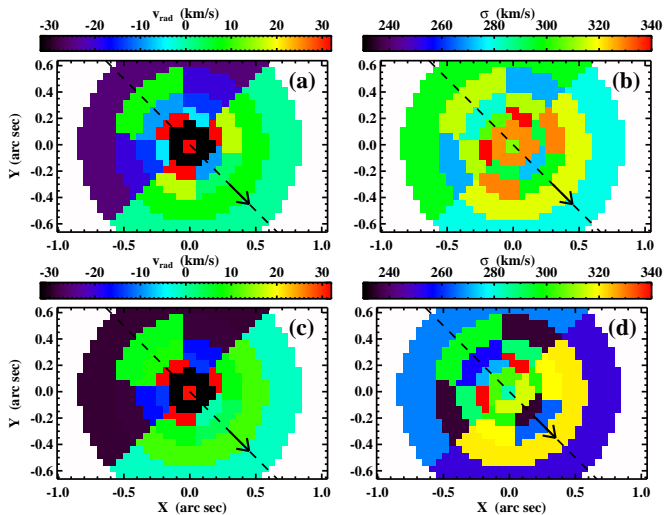,width=3.5in}
 \caption{Comparison of Gaussian and non-parametric LOSVD extraction methods.  \textbf{(a)} Radial velocity map of NGC 6086 from OSIRIS, derived by fitting Gaussian profiles to non-parametric LOSVDs.  \textbf{(b)} Velocity dispersion map from fitting Gaussian profiles to non-parametric LOSVDs.  \textbf{(c)} Radial velocity map, derived by fitting Gaussian LOSVDs to OSIRIS spectra.  \textbf{(d)} Velocity dispersion map from fitting Gaussian LOSVDs to OSIRIS spectra.  The dashed line in each figure represents the major axis of the galaxy, with the arrow pointing north.  The median error values are (a) 41 km s$^{-1}$; (b) 26 km s$^{-1}$; (c) 35 km s$^{-1}$; and (d) 32 km s$^{-1}$.  Radial velocities and dispersions from the two extraction methods are consistent within errors.}
\label{fig:GHcompare}
\end{figure}

Uncertainties for each LOSVD are determined by 100 Monte-Carlo trials.  In each trial, random noise is added to the galaxy spectrum, according to the RMS value of the original fit, and the LOSVD fitting process is repeated.  At each velocity bin, the uncertainties $\sigma_{\mathcal{L}+}$ and $\sigma_{\mathcal{L}-}$ are computed from the distribution of trial LOSVD values.  We then adjust the uncertainties in the wings of each LOSVD, so that $\mathcal{L} - \sigma_{\mathcal{L}-} = 0$.  

Stellar template mismatch can be a major source of systematic error in determining LOSVDs \citep[e.g.,][]{Carter,SG03,Emsellem}.  To address this issue, we have directly observed a diverse set of late-type template stars.  Our nine templates from OSIRIS are giant, supergiant, and dwarf stars with spectral types from G8 to M4.  We have found the most appropriate template for OSIRIS spectra of NGC 6086 to be HD 110964, an M4III star.  In Appendix~\ref{app:psf}, we describe our method for choosing the optimal template star, and how fitting LOSVDs with different template stars influences measurements of $M_\bullet$ and M$_\star$/L$_R$.  We fit GMOS spectra of NGC 6086 with a single G9III template star, HD 73710, which we observed with GMOS.  The calcium triplet region is less sensitive to template mismatch than other optical and near-infrared regions used to measure kinematics \citep{Barth02}.  

\subsection{Two-Dimensional Kinematics in NGC 6086}
\label{sec:losvd}

Our integral-field observations uncover complex kinematic structures within the central 3.1 kpc ($4.9''$) of NGC 6086.  In Figure~\ref{fig:kinmap}, we display 2-dimensional maps of kinematic moments from OSIRIS and GMOS data; we have computed $v_{\rm rad}$, $\sigma$, $h_3$, and $h_4$ by fitting a 4th-order Gauss-Hermite polynomial to each non-parametric LOSVD.  For data with modest signal-to-noise, non-parametric LOSVDs must be used with caution, as noise may falsely introduce strong non-Gaussian components to the fit.  In Figure~\ref{fig:GHcompare}, we compare two estimates of $v_{\rm rad}$ and $\sigma$ from OSIRIS spectra.  One estimate is obtained by fitting Gaussian profiles to non-parametric LOSVDs (Figures 7a and 7b), and the other is obtained by forcing a Gaussian LOSVD to fit the original spectra (Figures 7c and 7d).  In every spatial region, $v_{\rm rad}$ and $\sigma$ from the two fitting options are consistent within errors, and so we can trust the non-parametric LOSVDs.  GMOS data show similar agreement between the two estimates.

The stellar velocity dispersions measured by OSIRIS and GMOS each peak within 250 pc of the galaxy center, but not at the central spatial bin.  Central decreases in velocity dispersion have been observed in several other early-type galaxies with known black holes \citep[e.g.,][]{vdM94,Pinkney,Geb07,Nowak08}. 
Possible physical explanations include an unresolved stellar disk or a localized population of young stars.  No dust features are present in photometry of NGC 6086, nor is there any evidence of an active galactic nucleus.  The radial velocities are highly disturbed within the central 200 pc, which are only resolved by OSIRIS: at maximum, $\Delta v_{\rm rad} = 194 \pm 52$ km s$^{-1}$.  \citet{Geb07} found similar patterns in $v_{\rm rad}$ and $\sigma$ in the central 100 pc of NGC 1399, which were reproduced by models with a high prevalence of tangential orbits.  Likewwise, our best-fitting model of NGC 6086 is tangentially biased in the central 200 pc; the average value of $\sigma_r / \sigma_t$ is 0.55.  However, the 2-dimensional structure of $v_{\rm rad}$ is not consistent with a resolved stellar disk.  Axisymmetric modeling of elliptical galaxies by \citet{Geb03}, \citet{Shapiro}, and \citet{SG10} suggests that tangential bias is common within the black hole sphere of influence.  

Figures 6c, 6d, 6g, and 6h illustrate the 2-dimensional behavior of the third- and fourth-order Gauss-Hermite moments, $h_3$ and $h_4$.  Within errors, our measurements are largely consistent with $h_3 = 0$, while $h_4$ is significantly negative, corresponding to LOSVDs with ``boxy''  shapes and truncated wings.

\begin{figure*}[htbp]
 \centering
  \epsfig{figure=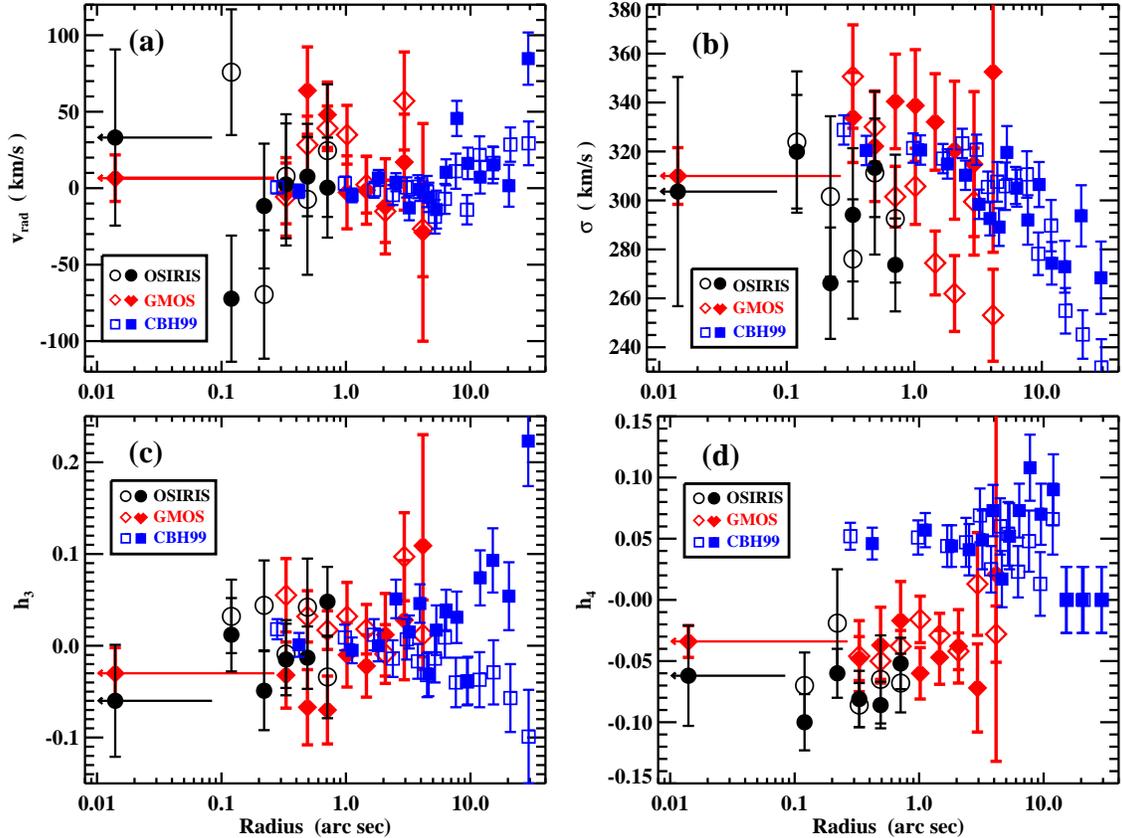,width=6.0in}
 \caption{Extracted kinematics along the major axis of NGC 6086.  Black circles are OSIRIS measurements, red diamonds are GMOS measurements, and blue squares are measurements from \citet{Carter}.  Filled symbols represent points from the positive (north) side of the major axis, and open symbols represent points from the negative (south) side.  For OSIRIS and GMOS data, the kinematic moments are computed from non-parametric LOSVDs.  \textbf{(a)}  Radial velocity, relative to the central velocity of NGC 6086.  The central velocity is defined separately for each dataset, such that the spatial averages to either side of the major axis are symmetric about $v_{\rm rad} = 0$.  Velocities from the negative (south) side of the major axis have been inverted.  \textbf{(b)}  Line-of-sight velocity dispersion.  \textbf{(c)}  Gauss-Hermite $h_3$, with values from the negative (south) side of the major axis inverted.  \textbf{(d)}  Gauss-Hermite $h_4$.}
\label{fig:kin}
\end{figure*}

We use major-axis kinematics from \citet{Carter} to constrain stellar orbit models at radii out to 18.9 kpc ($29.3''$), several times the extent of our IFS data.  To incorporate these data into our models, we have adopted higher uncertainties than the values quoted in \citet{Carter}; our treatment attempts to account for additional systematic errors, which are described by \citet{Carter} but excluded from their published measurements for NGC 6086.  In Figure~\ref{fig:kin}, we compare kinematic moments from \citet{Carter} to the moments derived from OSIRIS and GMOS, selecting the spatial bins along the galaxy's major axis.  We invert the sign of $v_{\rm rad}$ and $h_3$ for bins on the southern half of the galaxy.  Values of $v_{\rm rad}$, $\sigma$, and $h_3$ measured from the three data sets largely agree, although the OSIRIS data yield somewhat smaller values of $\sigma$.  At radii between $0.6''$ and $4.9''$, GMOS spectra from the southern half of the galaxy yield significantly lower values of $\sigma$ than spectra from the northern half.  The average asymmetry is 35 km s$^{-1}$; our median error for individual GMOS measurements is 16 km s$^{-1}$.  The asymmetry is not seen in long-slit data, which more consistently agree with GMOS along the north side of the major axis.  In spatial bins corresponding to the minor axis, $v_{\rm rad}$ and $\sigma$ behave similarly to the major-axis trends depicted in Figure~\ref{fig:kin}, and agree with minor-axis kinematics from \citet{Loubser}.  Beyond the central 200 pc, we find no convincing signs of kinematically distinct stellar populations dominating galaxy spectra at 0.5, 0.9, and 1.6 $\mu$m.

The most significant discrepancy in the major-axis kinematics is between the negative values of $h_4$ derived from IFS data and the positive values of $h_4$ measured by \citet{Carter} (Figure 8d).  Given the uniformity of the positive and negative values over many radial bins, we attribute this discrepancy to systematic errors in at least one set of measurements.  As noted by \citet{Carter} (and references therein), stellar template mismatch can bias $h_4$; however, this study and \citet{Carter} both perform careful analysis with multiple stellar templates (9 and 28 stars, respectively).  \citet{Nowak08} demonstrated that large spectral smoothing factors can also bias $h_4$ to negative values.  Still, a smoothing factor $> 100$ would be necessary to produce the full discrepancy between our values and those from \citet{Carter}.  Another source of error could be adjustments to the equivalent widths of absorption features in galaxy and template spectra; we address this issue in Appendix~\ref{app:psf}.  Regardless of the cause, including discrepant data in stellar orbit models can influence the best-fit solutions.  We discuss the effects on measurements of $M_\bullet$ and M$_\star$/L$_R$ in Section~\ref{sec:systerr}. 

%
%

%
\section{Stellar Orbit Models and Black Hole Mass}
\label{sec:results}

\subsection{Stellar Orbits}
\label{sec:model}

We generate stellar orbit models of NGC 6086, using the static potential method introduced by
\citet{Schild}.  We use the axisymmetric modeling algorithm described in detail in
Gebhardt et al. (2000b; 2003), Thomas et al. (2004; 2005), and \citet{Siopis}.
Here we provide a summary of the procedure.  Similar models are presented in \citet{RT84}, \citet{Rix97}, \citet{Cretton}, and \citet{VME04}.

We assume that the central region of NGC 6086 consists of three mass
components -- stars, a central black hole, and an extended dark matter halo
-- described by the radial density profile
\begin{equation}
  \rho(r) = \frac{M_\star}{L_R} \nu(r) +  M_\bullet \delta(r) + \rho_{\rm halo}(r) \,.
\label{eq:rho}
\end{equation}
The stellar distribution is assumed to follow the observed (deprojected) luminosity
density $\nu(r)$ (see Figure~\ref{fig:lden}) with a constant stellar mass-to-light ratio M$_\star$/L$_R$.
For the dark matter halo, we compare two density profiles: the commonly used 
NFW form \citep{NFW}, and 
a logarithmic (LOG) profile\footnote{The gravitational potential is logarithmic: $\Phi = \frac{1}{2} v_c^2$ ln($r^2 + r_c^2$).}:
\begin{equation}  
    \rho_{\rm halo}(r) = \frac{v_c^2}{4\pi G} \frac{3r_c^2 + r^2}{\left( r_c^2 + r^2 \right) ^2} \,.
\label{eq:halo}
\end{equation}
The free parameters in the LOG profile are the asymptotic circular speed $v_c$ and the core radius $r_c$, within which the density is approximately constant.  The enclosed halo mass for this profile,
\begin{equation}  
   M_{\rm halo}(<r) = \frac{v_c^2 \, r}{G} \left( 1 - \frac{r_c^2}{r_c^2 + r^2} \right) \; ,
\label{eq:menc}
\end{equation}
is predominantly set by $v_c$.  The difference between the NFW and LOG
profiles is greatest at small radii, where the NFW profile yields higher
densities, $\rho_{\rm halo} \propto r^{-1}$.  However, each profile is greatly exceeded
by the stellar mass density in the inner regions of NGC 6086.  
As described below, we have
compared LOG and NFW profiles in a subset of models, and find no
significant differences in the best-fit values of $M_\bullet$.

For a given set of input parameters $M_\bullet$, M$_\star$/L$_R$, and $\rho_{\rm halo}$, we
compute a continuous, static gravitational potential
from Equation~(\ref{eq:rho}).  Azimuthal symmetry about the $z$-axis
(corresponding to the projected minor axis) is imposed, as well as symmetry
about the equatorial plane ($z = 0$).
We then generate stellar orbits by propagating test particles through the potential.  Orbits are tracked in a finely spaced polar grid, $(r,\theta)$, where $\theta$ is the polar angle from the $z$-axis.  Our models of NGC 6086 use 96 radial and 20 polar bins per quadrant.  Each orbit is sampled at a random set of azimuthal angles, $\phi$.

The initial phase space coordinates of test particles are chosen to sample
thoroughly three integrals of motion: energy $E$, angular momentum
component $L_z$, and the third, non-classical integral, $I_3$.
Computational noise and finite propagation steps introduce noise into test
particle trajectories; this is mitigated by allowing each particle to
complete 200 circuits of the potential and then determining its average
orbit.  Orbits that escape the potential are not included in subsequent
fitting.  For a given potential, our model of NGC 6086 includes
approximately $16,000$ to $19,000$ bound orbits.  Identical counterparts
with the opposite sign of $L_z$ raise the total to 32,000 - 38,000 orbits.
Each orbit in the model is assigned a scalar weight; initially, all bound
orbits are given equal weights.  

The set of best-fit orbital weights is determined by comparing projected LOSVDs from the orbits to the observed LOSVDs for the galaxy.  Our models use non-parametric LOSVDs, defined in 15 velocity bins between $-1000$ and 1000 km s$^{-1}$.  Each observed LOSVD spatially maps to a linear combination of bins within the model, according to the spatial boundaries of the corresponding spectrum, and to the instrument-specific PSF.  A corresponding model LOSVD is computed from the projected velocity distributions of individual orbits in each spatial bin, the appropriate combination of spatial bins, and the orbital weights.  Only the orbital weights are varied to determine the best-fit solution.

The best-fit solution is determined by the method of maximum entropy, as in \citet{RT88}.  This method maximizes the function $f \equiv S - \alpha\chi^2$, where
\begin{equation} \chi^2 = \sum_i^{N_b} \sum_j \frac{\left[ \mathcal{L}_{i, \rm data}\left( v_j \right) - \mathcal{L}_{i, \rm model}\left( v_j \right) \right] ^2 }{\sigma^2_i \left( v_j \right)}
\label{eq:chi2}
\end{equation}
and
\begin{equation} S = - \sum_k w_k \, {\rm ln}  \left( \frac{w_k}{V_k} \right) 
\label{eq:entropy}
\end{equation}
Here, $\mathcal{L}_{i,\rm data}$ and $\mathcal{L}_{i,\rm model}$ are LOSVDs in each of the $i = 1, \, ... \, N_b$ spatial bins, $\sigma^2_i$($v_j$) is the squared uncertainty in $\mathcal{L}_{i,\rm data}$ at velocity bin $v_j$, $w_k$ is the orbital weight for the $k$th orbit, and $V_k$ is the phase volume of the $k$th orbit.  The parameter $\alpha$ is initially small so as to distribute orbital weights broadly over phase space, and is increased over successive iterations so that the final optimization steps exclusively minimize $\chi^2$.  A further constraint for all solutions is that the summed spatial distribution of all weighted orbits must match the observed luminosity density profile. 

%
\begin{figure*}[htbp]
 \centering
  \epsfig{figure=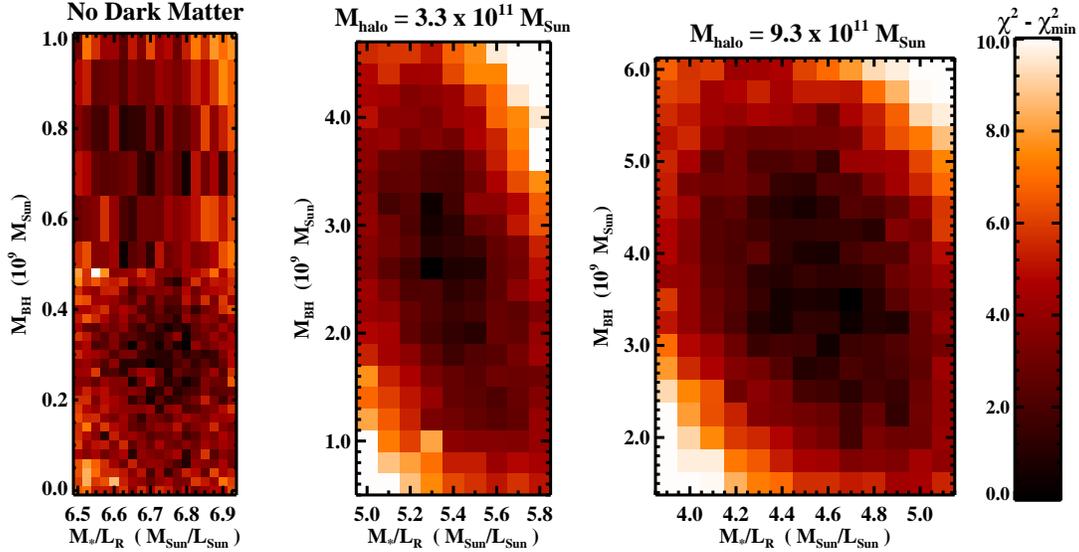,width=5.8in}
 \caption{Surface plots of $\chi^2 \; vs.$ M$_\star$/L$_R$ and $M_\bullet$, for models fitting both OSIRIS and GMOS data.  \textit{Left:} Models with no dark matter halo.  Long vertical pixels represent regions with coarser sampling in $M_\bullet$.  \textit{Middle:} $v_c = 300$ km s$^{-1}$ and $r_c = 8.0$ kpc, for $M_{\rm halo} = 3.3 \times 10^{11}$ M$_\odot$ within 18.9 kpc.   \textit{Right:} $v_c = 500$ km s$^{-1}$ and $r_c = 8.0$ kpc, for $M_{\rm halo} = 9.3 \times 10^{11}$ M$_\odot$ within 18.9 kpc.  For each dark matter halo, additional models were run outside the range of M$_\star$/L$_R$ and $M_\bullet$ depicted here.  These models all yield higher values of $\chi^2$.}
\label{fig:chsqsurf}
\end{figure*}
%

%
\begin{figure}[htbp]
 \centering
       \epsfig{figure=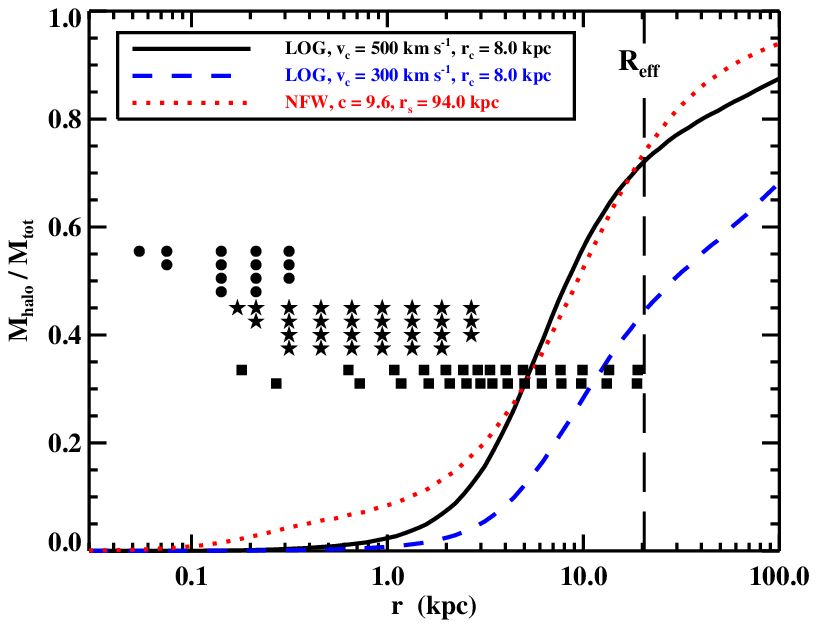,width=3.5in}
 \caption{Dark matter fraction of total enclosed mass, for the model dark matter halos presented in Table~\ref{tab:allres}.  For each halo, the total enclosed mass includes the best-fit black hole mass and best-fit stellar mass-to-light ratio, derived using OSIRIS, GMOS, and long-slit data.  The dashed vertical line marks the effective radius derived from our photometric data.  Filled symbols represent the radial positions of spectra from OSIRIS (circles), GMOS (stars), and \citet{Carter} (squares); the vertical positions of the symbols have no physical meaning.}
\label{fig:DMfraction}
\end{figure}

\subsection{Black Hole Mass and Mass-to-Light Ratio}
\label{sec:BHres}

Our combination of IFS and long-slit data within $30''$ is not sufficient
to measure directly the dark matter halo profile or enclosed mass.  We
therefore consider different halo masses and profiles in our analysis.  For a given dark
matter halo profile, we generate a set of 32,000-38,000 stellar orbits for an
input $M_\bullet$ and M$_\star$/L$_R$, and obtain $\chi^2$ for the best-fit
orbital weights using the method described in the previous subsection.
This process is repeated over a finely sampled grid in $M_\bullet$ and
$M_\star/L_R$.  We have completed all trials for NGC 6086 on supercomputers at the Texas Advanced Computing Center (TACC), totaling $\sim 10,000$ CPU hours.  Our results are summarized in Table~\ref{tab:allres}.  

%
\begin{table*}[htbp]
\begin{center}
\caption{Results from stellar orbit models.}
\label{tab:allres}
\begin{tabular}[b]{cccccccccc}  
\hline
Data & $v_c$ & $r_c$ & $c$ & $r_s$ & $M_{\rm halo}$ & $M_\bullet$  & M$_\star$/L$_R$ & $\chi^2_{\rm min}$ & $N_{\rm dof}$\\
& (km s$^{-1}$) & (kpc) & & (kpc) & ($10^{11} \; \rm M_\odot$) & ($10^9 \; \rm M_\odot$) & (M$_\odot$/L$_\odot$) &  & \\
\\
(1) & (2) & (3) & (4) & (5) & (6) & (7) & (8) & (9) & (10)\\
\hline 
\\
OSIRIS + GMOS + \citet{Carter} & 0.01 & 20.0 & & & $\approx 0$ & $0.6^{+0.4}_{-0.4} \; \left( 0.6^{+0.4}_{-0.4} \right)$ & $6.68^{+0.15}_{-0.17} \; \left( 6.7^{+0.2}_{-0.9} \right)$ & 1114.4 & 795\\
\\
OSIRIS + GMOS + \citet{Carter} & 300 & 8.0 & & & 3.34 & $2.6^{+1.0}_{-1.0} \; \left( 2.6^{+1.3}_{-1.0} \right)$ & $5.4^{+0.3}_{-0.3} \; \left( 5.4^{+0.3}_{-0.8} \right)$ & 1019.4 & 795\\
\\
OSIRIS + GMOS + \citet{Carter} & 500 & 8.0 & & & 9.28 & $3.6^{+1.2}_{-1.1} \; \left( 3.6^{+1.7}_{-1.1} \right)$ & $4.6^{+0.3}_{-0.4} \; \left( 4.6^{+0.3}_{-0.7} \right)$ & 1010.2 & 795\\
\\
OSIRIS + GMOS + \citet{Carter} & & & 9.6 & 94.0 & 9.05 & $3.6^{+1.1}_{-1.2} \; \left( 3.6^{+1.6}_{-1.2} \right)$ & $4.3^{+0.3}_{-0.4} \; \left( 4.3^{+0.3}_{-0.7} \right)$ & 1009.4 & 795\\
\\
OSIRIS + GMOS & 0.01 & 20.0 & & & $\approx 0$ & $3.2^{+1.5}_{-1.3} \; \left( 3.2^{+1.8}_{-1.3} \right)$ & $4.4^{+0.4}_{-0.3} \; \left( 4.4^{+0.4}_{-0.6} \right)$ & 622.9 & 345\\
\\
OSIRIS + GMOS & 500 & 8.0 & & & 9.28 & $3.5^{+1.4}_{-1.5} \; \left( 3.5^{+1.8}_{-1.5} \right)$ & $4.1^{+0.4}_{-0.5} \; \left( 4.1^{+0.4}_{-0.7} \right)$ & 623.8 & 345\\
\\
OSIRIS + \citet{Carter} & 500 & 8.0 & & & 9.28 & $1.9^{+1.4}_{-1.1} \; \left( 1.9^{+1.5}_{-1.1} \right)$ & $4.6^{+0.7}_{-0.7} \; \left( 4.6^{+0.7}_{-0.9} \right)$ & 412.7 & 570\\
\\
GMOS + \citet{Carter} & 500 & 8.0 & & & 9.28 & $7^{+2}_{-3} \; \left( 7^{+3}_{-3} \right)$ & $4.5^{+0.5}_{-0.5} \; \left( 4.5^{+0.5}_{-0.8} \right)$ & 725.5 & 675\\
\\
\hline
\end{tabular}
\end{center}
\begin{small}
\textbf{Notes:}  Column 1: Data sets included in trial.   Column 2: circular velocity of LOG dar matter halo (Eq.~\ref{eq:halo}).  Column 3: core radius of LOG dark matter halo (Eq.~\ref{eq:halo}).  Column 4: concentration parameter for NFW dark matter halo.  Column 5: scale radius for NFW dark matter halo.  Column 6: enclosed halo mass, defined at the outermost long-slit data point.  The corresponding radius is 18.9 kpc.  The ``no dark matter'' case has $v_c = 0.01$ km s$^{-1}$, $r_c = 20.0$ kpc, and $M_{\rm halo} \sim 200$ M$_\odot$.  Column 7: best-fit black hole mass.  Quoted errors correspond to 68\% confidence intervals.  Values in parentheses include all systematic errors.  Column 8: best-fit $R$-band stellar mass-to-light ratio.  Quoted errors correspond to 68\% confidence intervals.  Values in parentheses include all systematic errors.  Column 9: minimum $\chi^2$ value for all models.  Column 10: degrees of freedom in model fits to LOSVDs.  Computed values include a smoothing factor of 1 degree of freedom per 2 velocity bins for non-parametric LOSVDs from OSIRIS and GMOS.
\end{small}
\end{table*}

In Figure~\ref{fig:chsqsurf}, we illustrate how $\chi^2$ in the orbit models varies with $M_\bullet$ and
M$_\star$/L$_R$.  Three dark matter halo masses are shown: no
dark matter (left panel), an intermediate LOG halo with $v_c=300$ km s$^{-1}$ (middle),
and a maximal LOG halo with $v_c=500$ km s$^{-1}$ (right).  The latter is chosen to
approximate the measured line-of-sight velocity dispersion of 302 km s$^{-1}$ for
NGC 6086's host cluster \citep{Zab}, which corresponds to a full
3-dimensional velocity dispersion of 523 km s$^{-1}$.
We set the core radius in Equation~(\ref{eq:halo}) to be $r_c =8.0$ kpc,
reflecting the value of 8.2 kpc determined by \citet{Thom07} for NGC 4889,
the Coma BCG.  Our dynamical models are constrained within a radius
of 18.9 kpc, corresponding to outermost radius of $29.3''$ for
long-slit data in \citet{Carter}.  We also have run models with a single NFW dark matter profile, constructed to contain the same enclosed mass within 18.9 kpc as our most massive LOG halo.  In cosmological $N$-body simulations, the NFW scaling parameters $c$ and $r_s$ are correlated according to the relationship
\begin{equation}
r_s^3 = \left( \frac{3 \times 10^{13} \; \rm M_\odot}{200 \frac{4\pi}{3} \rho_{\rm crit} \, c^3} \right) 10^{\frac{1}{0.15} \left( 1.05 - \rm log_{10} \it c \right)}
\end{equation}
\citep{NFW,Rix97}\footnote{Note the erratum in Equation (B3) of \citet{Rix97}; the correct equation is log$_{10} c = 1.05 - 0.15$ log$_{10} \left( M_{200} / 3 \times 10^{13} \rm M_\odot \right)$.}, where $\rho_{\rm crit} = 3H_0^2/8\pi G$.
Combining this relationship with our enclosed mass constraint, we obtain $c  = 9.6$ and $r_s = 94.0$ kpc.  

Dark matter is ubiquitous in galaxies, thus motivating its inclusion in stellar orbit models.  Furthermore, models with dark matter produce better fits to our full set of kinematics: when the dark matter component is removed, $\chi^2_{min}$ increases by $\sim 100$ (Table~\ref{tab:allres}).  In Appendix~\ref{app:losvd} we compare our full sets of observed LOSVDs to the best-fitting models with and without dark matter.  The largest discrepancies between the two models occur at radii beyond $15''$, where we only have a few data points from \citet{Carter}.  Without thorough radial coverage or multiple long-slit position angles, we cannot fully untangle degeneracies between M$_\star$/L$_R$, $v_c$, and $r_c$ (or M$_\star$/L$_R$, $c$, and $r_s$ in the case of an NFW profile).

In Figure~\ref{fig:DMfraction} we display the dark matter fraction as a function of radius, for each of the halos described above.  In each case, we use the best-fit values of $M_\bullet$ and M$_\star$/L$_R$, described below, to compute the total enclosed mass.  Using our surface brightness profile from \textit{HST}/KPNO we compute an effective radius, $R_{\rm eff}$, of $31.7''$ (20.4 kpc), defined as the semi-major axis of the elliptical isophote containing half of the total luminosity.   We assume a total luminosity of $1.82 \times 10^{11}$ L$_{\odot,R}$, from $M_V = -23.11$ in \citet{Lauer07} and $V-R = 0.64$.  Within $R_{\rm eff}$, dark matter composes $44\%$ to $74\%$ of the total mass (for the intermediate-mass LOG halo and the NFW halo, respectively).  This range agrees with dynamical models of other cD galaxies with LOG and NFW halos: \citet{Thom07} found $\sim 50-75\%$ dark matter within $R_{\rm eff}$ for NGC 4889 and NGC 4874, and \citet{GT09} found $\sim 40\%$ dark matter within $R_{\rm eff}$ for M87.  Within the $4.9''$ outer radius of GMOS data, the maximum dark matter fraction in our models is $20\%$. 

\begin{figure}[htbp]
 \centering
       \epsfig{figure=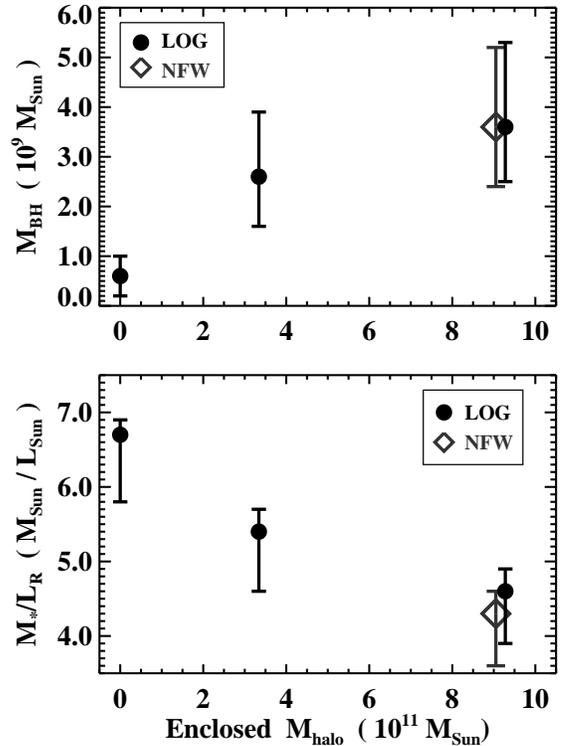,width=3.0in}
 \caption{\textit{Top:}  Best-fit black hole mass for different assumed dark matter halos.  \textit{Bottom:}  Best-fit stellar mass-to-light ratio.  Black circles represent LOG halos, and grey diamonds represent the NFW halo.  Each measurement comes from fitting the full set of LOSVDs from OSIRIS, GMOS, and \citet{Carter}.  The enclosed halo mass is defined at 18.9 kpc, the radius of the outermost data point from \citet{Carter}.  Error bars indicate 68\% confidence, and include systematic errors.}
\label{fig:halores}
\end{figure}

In Figure~\ref{fig:halores} we illustrate the variation of $M_\bullet$ and
M$_\star$/L$_R$ with enclosed halo mass.  We compute best-fit values for $M_\bullet$ and M$_\star$/L$_R$ by integrating the 2-dimensional likelihood function from each $\chi^2$ surface; we describe this method and our determination of errors in Section~\ref{sec:error}.  Figures~\ref{fig:chsqsurf} and~\ref{fig:halores} show that the best-fit
values of $M_\bullet$ and M$_\star$/L$_R$ are substantially influenced by
the presence of dark matter in the stellar orbit models.
This occurs because our innermost kinematics sample the nucleus of the
galaxy, where orbits are dominated by the enclosed mass of stars and the
central black hole, whereas enclosed stellar and dark halo masses are both
important at larger radii.  The significant presence of dark matter at
large radii drives the best-fit models to lower values of M$_\star$/L$_R$.
In turn, the decreased stellar mass requires a higher black hole mass to
reproduce the kinematics in the nucleus.  This trend was initially
demonstrated by \citet{GT09}, for M87.  Negative covariance between $M_\bullet$ and M$_\star$/L$_R$ is also visible in $\chi^2$ contours for individual dark matter halo models (Figure~\ref{fig:chsqsurf}).  In Section~\ref{sec:disc}, we compare our best-fit values of $M_\bullet$ from different dark matter halo models to the predictions from the $M_\bullet - \sigma$ and $M_\bullet - L$ relationships.

%

Using an NFW profile yields a $7\%$ decrease in the best-fit value of M$_\star$/L$_R$, relative to the LOG profile with the same enclosed mass at 18.9 kpc.  This is because the majority of our kinematic measurements occur at $r \leq 5$ kpc, where the more centrally-concentrated NFW profile yields higher dark matter densities.  In the central 100 pc, the stellar core of NGC 6086 varies nearly as $r^{-1}$ in luminosity density, mimicking the slope of the NFW profile.  Near the black hole, the lower stellar mass density balances the higher density in dark matter, and so the best-fit values of $M_\bullet$ are identical for the NFW and LOG profiles.  We find $\chi^2_{\rm min, \, LOG} - \chi^2_{\rm min, \, NFW} = 0.8$, indicating no significant difference in the goodness of fit.  \citet{Thom05,Thom07} and \citet{GT09} have found similar difficulties in distinguishing between NFW and LOG profiles.    

A second way to address the influences of dark matter on M$_\star$/L$_R$ and $M_\bullet$ is to fit the orbit models only at radii where dark matter composes a small fraction of the enclosed mass.  For NGC 6086, we have run two trials in which we only fit LOSVDs from OSIRIS and GMOS: one trial with the maximum-mass LOG dark matter halo described above, and one trial with no dark matter.  In both of these trials, the best-fit values of $M_\bullet$ and M$_\star$/L$_R$ agree with our results from fitting IFS and long-slit data with the maximum-mass LOG halo (see Table~\ref{tab:allres}).  This agreement provides strong evidence that $M_\bullet \sim 3 \times 10^9$ M$_\odot$, regardless of our insensitivity to the exact structure of the dark matter halo in NGC 6086.  In contrast, forcing models without dark matter to fit long-slit data biases the best-fit black hole mass to a substantially lower value, $\sim 6 \times 10^8$ M$_\odot$, and increases the best-fit stellar mass-to-light ratio to 6.7 M$_\odot$/L$_\odot$.  This mass-to-light ratio is highly inconsistent with stellar population estimates, as we discuss in Section~\ref{sec:disc}.  Even though excluding long-slit data results in consistency between models with and without dark matter, these models are not as thoroughly constrained, and we obtain slightly larger confidence intervals in $M_\bullet$ and M$_\star$/L$_R$ for each trial.

\begin{figure*}[htbp]
 \centering
  \epsfig{figure=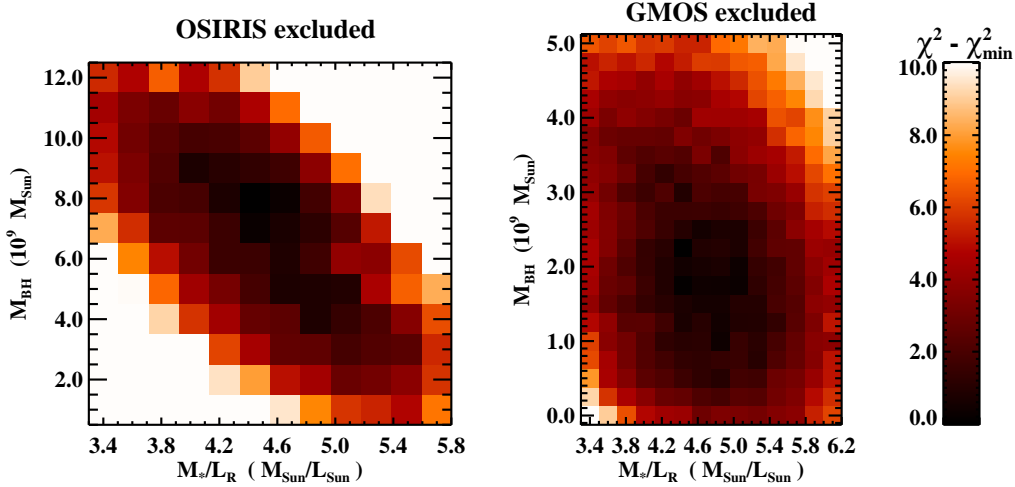,width=5.5in}
 \caption{Surface plots of $\chi^2 \; vs.$ M$_\star$/L$_R$ and $M_\bullet$, using integral-field data from different instruments.  \textit{Left:} GMOS data plus long-slit data from \citet{Carter}.  \textit{Right:} OSIRIS data plus long-slit data from \citet{Carter}.  All models include a LOG dark matter halo with $v_c = 500$ km s$^{-1}$ and $r_c = 8.0$ kpc, yielding $M_{\rm halo} = 9.3 \times 10^{11}$ M$_\odot$ within 18.9 kpc.  In each case, additional models were run outside the range of M$_\star$/L$_R$ and $M_\bullet$ depicted here.  These models all yield higher values of $\chi^2$.}
\label{fig:OGsurf}
\end{figure*}

Our data from OSIRIS, GMOS, and \citet{Carter} play complementary roles in constraining the gravitational potential of NGC 6086.  In Figure~\ref{fig:OGsurf}, we compare model results for one dark matter halo (LOG; $v_c = 500$ km s$^{-1}$), using data only from GMOS and \citet{Carter}, versus only from OSIRIS and \citet{Carter}.  The GMOS data are sufficient to detect a black hole, in part because of excellent seeing.  Yet the strong diagonal contours in the left panel of Figure~\ref{fig:OGsurf} indicate that the black hole mass derived from GMOS is degenerate with the enclosed stellar mass.  LOSVDs from OSIRIS have large statistical errors and by themselves cannot place strong constraints on the black hole mass.  However, the OSIRIS data help separate the respective influences of the stars and the black hole.  Using GMOS and OSIRIS data together reduces covariance between $M_\bullet$ and M$_\star$/L$_R$ and lowers the statistical uncertainties of both quantities.  Long-slit data from \citet{Carter} confirm the presence of dark matter and tighten constraints on M$_\star$/L$_R$ and $M_\bullet$ for individual dark matter halo models.

%

It is not clear how to interpret the significant increase in the best-fit value of $M_\bullet$, from $1.9^{+1.5}_{-1.1} \times 10^9$ M$_\odot$ with OSIRIS and long-slit data only, to $7^{+3}_{-3} \times 10^9$ M$_\odot$ with GMOS and long-slit data only.  Using central $\sigma = 329$ km s$^{-1}$ from \citet{Loubser}, and $M_\bullet = 3.6 \times 10^9$ M$_\odot$ from our combined-data trial, we compute $r_{\rm inf} = 0.22''$.  In this case, GMOS marginally resolves the sphere of influence, with a seeing FWHM $\sim 2 r_{\rm inf}$.  At small radii, LOSVDs from GMOS yield slightly higher velocity dispersions than overlapping LOSVDs from OSIRIS (Figure 8b), which could contribute to the increase in $M_\bullet$.  With no obvious way to assess independently the accuracy of data from OSIRIS versus GMOS, we favor including both sets of LOSVDs.  The corresponding black hole mass is $3.6^{+1.7}_{-1.1} \times 10^9$ M$_\odot$, which lies between the two partial-data values and has the narrowest confidence interval.  The confidence interval in M$_\star$/L$_R$ is also minimized by including all data, though the best-fit value of $4.6^{+0.3}_{-0.7}$ does not change significantly upon exclusion of OSIRIS or GMOS data.  

None of the $\chi^2$ surfaces in Figures~\ref{fig:chsqsurf} and~\ref{fig:OGsurf} are completely smooth.  In particular, the models without any dark matter show large variations in $\chi^2$ over small changes in $M_\bullet$ and M$_\star$/L$_R$ (Figure~\ref{fig:chsqsurf}).  We suspect that these variations arise from numerical noise in propagating test particles through different potentials: each small change in the potential may send a given test particle through a different set of spatial regions.  The cumulative effect is that each model creates a different set of test-particle LOSVDs, and $\chi^2$ can change abruptly in spite of the freedom to adjust orbital weights.  When the models include a constant dark matter component, the relative changes in the potential are smaller, and the noise in $\chi^2$ is less pronounced.  Still, the $\chi^2$ surface for each dark matter halo exhibits a noise floor at the level of $\Delta\chi^2 \sim 1$.  In Section~\ref{sec:conf}, we describe how this noise influences our measurements of confidence intervals.

\subsection{Determining Errors}
\label{sec:error}

The figure of merit for evaluating confidence intervals in $M_\bullet$ and M$_\star$/L$_R$ is $\Delta\chi^2 \equiv \chi^2 - \chi^2_{\rm min}$,
where $\chi^2_{\rm min}$ is the lowest output value among all models.  For NGC
6086, we determined confidence intervals by integrating the relative likelihood function, $P \propto
e^{-\frac{1}{2}\Delta\chi^2}$.  Although $\Delta\chi^2$ is a better statistical indicator than $\chi^2$ per degree of freedom \citep[e.g.,][]{vdM98,Geb03}, the latter is useful for crudely indicating the level of agreement between the data and the model with the best fit.  For each model, the number of degrees of freedom, $N_{\rm dof}$, depends on the number of observed LOSVDs and the number of velocity bins evaluated per LOSVD.  Because we used a spectral smoothing factor in determining LOSVDs for OSIRIS and GMOS, the velocity bins are not entirely independent.  We estimate that each LOSVD from OSIRIS or GMOS has 1 degree of freedom per 2 velocity bins, whereas long-slit data from \citet{Carter} has 1 degree of freedom per velocity bin.  For all experiments with NGC 6086, we find $\chi^2_{\rm min} / N_{\rm dof}$ between 0.7 and 1.8, indicating reasonable agreement.

\subsubsection{Confidence Intervals}
\label{sec:conf}

We determine confidence intervals by using $\Delta\chi^2$ as an empirical measure of relative likelihood between models with different $M_\bullet$ and M$_\star$/L$_R$, and by numerically integrating this likelihood with respect to $M_\bullet$ and M$_\star$/L$_R$.  In contrast to the majority of previous studies 
(e.g., Gebhardt et al. 2000b; 2003; 2007; Nowak et al. 2007; 2008; Gebhardt \& Thomas 2009; G\"{u}ltekin et al. 2009b; Siopis et al. 2009; Shen \& Gebhardt 2010; cf. van der Marel et al. 1998), 
we do not use fixed values of $\Delta\chi^2$ to define confidence intervals.  The fixed $\Delta\chi^2$ method is appropriate only if the orbit models cleanly sample a well-defined likelihood function of $M_\bullet$ and M$_\star$/L$_R$; 
other studies typically assume a 2-dimensional Gaussian likelihood function. 
Our models of NGC 6086 produce noisy $\chi^2$ contours (Figure~\ref{fig:chsqsurf}), in which case the fixed $\Delta\chi^2$ method is sensitive to noise in individual models.  This effect is especially pronounced for models without a dark matter halo.

\begin{figure*}[htbp]
 \centering
  \epsfig{figure=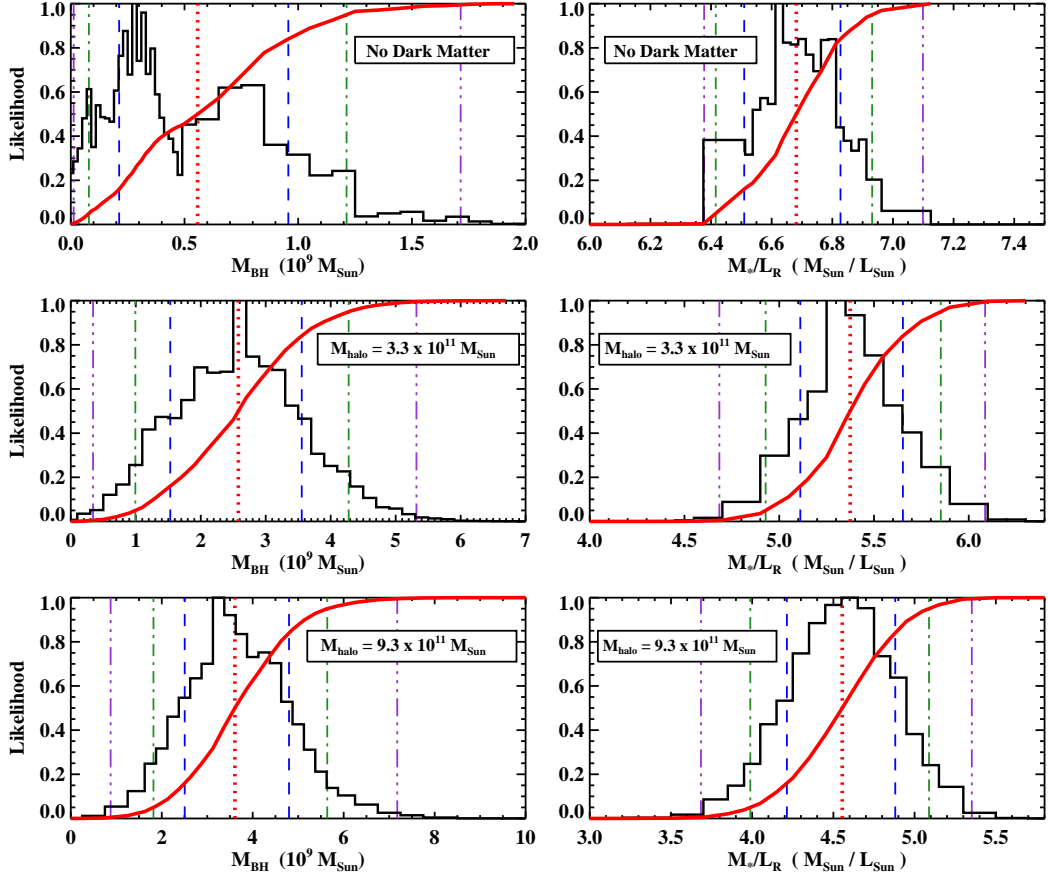,width=5.6in}
 \caption{Likelihood functions $P(M_\bullet)$ and $P($M$_\star$/L$_R)$, after marginalizing over one variable (M$_\star$/L$_R$ and $M_\bullet$, respectively) and re-normalizing.  The solid red line in each plot is the cumulative distribution.  Vertical lines in each figure represent the median value (dotted red line) and intervals for 68\%, 90\%, and 99\% confidence (dashed lines).  \textit{Top:} Models with no dark matter halo.  \textit{Middle:} $v_c = 300$ km s$^{-1}$ and $r_c = 8.0$ kpc, for $M_{\rm halo} = 3.3 \times 10^{11}$ M$_\odot$ within 18.9 kpc.   \textit{Bottom:} $v_c = 500$ km s$^{-1}$ and $r_c = 8.0$ kpc, for $M_{\rm halo} = 9.3 \times 10^{11}$ M$_\odot$ within 18.9 kpc.}
\label{fig:lik}
\end{figure*}

We define likelihood $P$ such that two models with $\chi^2_1$ and $\chi^2_2$ have relative likelihood
\begin{equation}  \frac{P_1}{P_2} = e^{-\frac{1}{2} \left( \chi^2_1 - \chi^2_2 \right)}
\label{eq:likrel}
\end{equation}
This form of $P$ is valid, provided that $\chi^2$ is measured from independent, Gaussian-distributed data points \citep{Cowan}; in our case, these are the observed LOSVDs.  To evaluate likelihood with respect to a single variable (i.e. $x \equiv M_\bullet$), we marginalize the 2-dimensional surface with respect to the other variable ($y \equiv$ M$_\star$/L$_R$), such that:
\begin{equation} P \left( x \right) \propto  \sum_{y_{\rm min}}^{y_{\rm max}} e^{-\frac{1}{2} \chi^2 \left( x, y \right)} \; \delta y \; , 
\label{eq:lik1d}
\end{equation}
where $\delta y$ is the interval between sampled values of $y$.  Confidence intervals in $x$ are determined by evaluating the cumulative distribution:
\begin{equation}  C \left( x \right) =  \frac{ \int_{x_{\rm min}}^x P \left( x' \right) dx' } {\int_{x_{\rm min}}^{x_{\rm max}} P \left( x' \right) dx'}
\label{eq:conf}
\end{equation}
In practice, we define [$x_{\rm min}$, $x_{\rm max}$] and [$y_{\rm min}$, $y_{\rm max}$] by expanding our range of models until the marginalized likelihood functions $P(M_\bullet)$ and $P$(M$_\star$/L$_R$) are nearly zero at the minimum and maximum modeled values of $M_\bullet$ and M$_\star$/L$_R$.  The physical limit $M_\bullet = 0$ is included in all trials.  For confidence level $k$, we define confidence limits at $C =  \frac{1}{2} \left( 1 \pm k \right)$.  For example, the 68\% confidence interval comprises all $x$ for which $0.16 \leq C \leq 0.84$.  For each trial, we define the best-fit values of $M_\bullet$ and M$_\star$/L$_R$ as the median values from $P(M_\bullet)$ and $P$(M$_\star$/L$_R$), corresponding to $C = \frac{1}{2}$.  In Figure~\ref{fig:lik}, we show $P(M_\bullet)$ and $P$(M$_\star$/L$_R$) for each dark halo setting, along with cumulative distributions, median values, and confidence intervals.  To estimate precise confidence limits, we linearly interpolate $C$ between discretely sampled values of $M_\bullet$ and M$_\star$/L$_R$.

%

Our empirical treatment yields wider 68\% confidence intervals than those derived from fixed $\Delta\chi^2$ and a Gaussian likelihood function.  Our intervals for confidence levels $\geq 90\%$ typically fall near those derived from fixed $\Delta\chi^2$.  By construction, our confidence intervals do not include $M_\bullet = 0$.  For the maximum-mass LOG dark matter halo in NGC 6086, the marginalized likelihood corresponding to $M_\bullet = 0$, $P(M_\bullet = 0)$, is $0.06\%$ of the maximum marginalized likelihood value.  
For a Gaussian likelihood function, this likelihood ratio would indicate $> 99.98\%$ confidence
for a black hole detection.  For models without dark matter, $P(M_\bullet = 0)$ is $23.2\%$ of the maximum value, and the detection falls to 91\% confidence.

An alternative way to determine confidence intervals from noisy $\chi^2$ is the method of \citet{vdM98}, in which random noise is added to the LOSVDs output by each orbit model, the $\chi^2$ surface is re-computed, and confidence limits are determined using fixed $\Delta\chi^2$.  This process is repeated in a Monte Carlo fashion, and the extrema of the confidence limits from all trials are adopted.  This treatment assumes that numeric noise in orbit models produces fluctuations about an intrinsically Gaussian likelihood distribution.  The global likelihood function of our models, however, is visibly non-Gaussian (see Figure~\ref{fig:lik}, top, in particular).

\subsubsection{Systematic Errors}
\label{sec:systerr}

The confidence intervals measured from $\chi^2$ surfaces account for statistical errors in the observed LOSVDs and random noise within the stellar orbit models.  Systematic errors must be addressed separately.  We have directly tested several systematic effects.

Our largest systematic error arises from discrepancies between LOSVD shapes derived from IFS versus long-slit data, as indicated by the parameter $h_4$ (Figure 8d).  To test how this discrepancy biases $M_\bullet$ and M$_\star$/L$_R$, we fit Gaussian profiles to the original LOSVDs from \citet{Carter}, constructing an alternative set of LOSVDs with $h_3 = 0$ and $h_4 = 0$.  Using these new LOSVDs in combination with the OSIRIS and GMOS data, we repeated the stellar orbit models for the maximum dark matter halo, and found $M_\bullet = 4.5^{+1.4}_{-1.1} \times 10^9$ M$_\odot$, and M$_\star$/L$_R = 4.2^{+0.3}_{-0.3}$ M$_\odot$/L$_\odot$.  This is a 22\% increase in $M_\bullet$ and a 9\% decrease in M$_\star$/L$_R$, relative to the corresponding trial with original LOSVDs from \citet{Carter}.  The direction of the bias indicates that more enclosed mass is required to produce LOSVDs with extended wings ($h_4 > 0$) at $\sim 3-19$ kpc.  Since the additional mass is obtained by increasing M$_\star$/L$_R$, the innermost LOSVDs drive the best fit toward a smaller value of $M_\bullet$.

Two additional sources of systematic error are the uncertainties of the optimal stellar template for fitting LOSVDs, and the average PSF for OSIRIS.  We find that both effects yield small errors.  Trials with 3 PSFs yield a standard deviation of 8\% in $M_\bullet$ and 1\% in M$_\star$/L$_R$.  Trials with 2 stellar templates differ by 5\% in $M_\bullet$ (corresponding to 3.6\% standard deviation) and 2\% in M$_\star$/L$_R$ (1.6\% standard deviation).  We describe these tests in detail in Appendix~\ref{app:psf}.

We use the following prescription to compute the total error in $M_\bullet$ for a given dark halo:
\begin{equation}  \sigma_{+, \, tot} =  \left( \sigma_{+, \, \chi^2}^2 + \sigma_{\rm PSF}^2 + \sigma_{\rm temp}^2 + 2\delta_{h_4}^2  \right) ^ {\frac{1}{2}}
\label{eq:uperr}
\end{equation}
and
\begin{equation}  \sigma_{-, \, tot} =  \left( \sigma_{-, \, \chi^2}^2 + \sigma_{\rm PSF}^2 + \sigma_{\rm temp}^2 \right) ^ {\frac{1}{2}} \; .
\label{eq:lowerr}
\end{equation}
Here, $\sigma_{+, \, tot}$ and $\sigma_{-, \, tot}$ are the upper and lower portions of the 68\% confidence interval, including all errors; $\sigma_{+, \, \chi^2}$ and $\sigma_{-, \, \chi^2}$ are the contributions from integrating the empirical likelihood function as in Section~\ref{sec:conf}, and represent statistical errors; $\sigma_{\rm PSF}$ is the standard deviation in best-fit $M_\bullet$ from trails with different OSIRIS PSFs; $\sigma_{\rm temp}$ is the standard deviation from trials with different template stars; and $\delta_{h_4}$ is the difference in best-fit $M_\bullet$ from using re-fit versus original LOSVDs from \citet{Carter}.  As excluding $h_3$ and $h_4$ from these LOSVDs introduces a bias toward higher $M_\bullet$, we assign this effect solely to $\sigma_{+, \, tot}$, with a 
magnitude of $2 \times \frac{1}{\sqrt{2}}\delta_{h_4}$.  
Our equations defining $\sigma_{+, \, tot}$ and $\sigma_{-, \, tot}$ for M$_\star$/L$_R$ are similar to 
(\ref{eq:uperr}) and (\ref{eq:lowerr}),
except we apply $\delta_{h_4}$ entirely to $\sigma_{-, \, tot}$.  To apply our results to different dark matter halo settings, we define the systematic errors as percentages, such that $\delta_{h_4}$, $\sigma_{\rm PSF}$ and $\sigma_{\rm temp}$ scale with $M_\bullet$ and M$_\star$/L$_R$.  We list the 68\% confidence intervals, including all errors, inside parentheses in Table~\ref{tab:allres}.  The confidence intervals outside parentheses in Table~\ref{tab:allres} only include $\sigma^2_{\pm, \, \chi^2}$.

By adding in quadrature, we have assumed zero correlation between different sources of systematic error; this is the most conservative approximation.  $\delta_{h_4}$ and $\sigma_{\rm temp}$ are likely correlated \citep[see, e.g.,][]{Carter}, but the contributions from $\sigma_{\rm temp}$ are small, and we have not run extensive tests to measure covariance between stellar templates and overall trends in $h_4$.  There is no obvious reason to suspect covariance between other terms.  

\subsubsection{Other Potential Sources of Error}
\label{sec:moreerror}

The shapes and depths of CO bandheads depend on luminosity class as well as spectral type \citep[e.g.,][]{SG03}.  Although we have extensively examined stars of multiple spectral types, our library of template stars from OSIRIS lacks an M-dwarf template; it is uncertain whether dwarf stars contribute significantly to spectra of NGC 6086.
Our sample of templates is also limited by the range of elemental abundances found in bright stars in the solar neighborhood.  \citet{Loub09} determine stellar metallicity at the center of NGC 6086 to be $\sim 2$ times solar ([Z/H] = $0.28 \pm 0.07$), with an $\alpha$-enhancement ratio, [E/Fe], of $0.39 \pm 0.04$.  Other BCGs are similarly metal- and $\alpha$-rich \citep{Brough07,Loub09}.  If LOSVDs derived from CaII, CO, and OH absorption features are sensitive to template star metallicities and $\alpha$ ratios, then our systematic error could be higher than estimated above.  

Another issue is the possibility of significant spatial variations in the stellar mass-to-light ratio, contradicting the uniformity imposed upon stellar obit models.  Radial gradients in age, metallicity, and $\alpha$-enhancement have been measured in individual BCGs \citep{Brough07,CGA10}; further modeling is necessary to quantify corresponding gradients in M$_\star$/L.

An untested source of systematic error for NGC 6086 is the assumed shape and inclination of the galaxy.  Our orbit models use an oblate, axisymmetric potential and assume edge-on inclination.
Previous studies have indicated that uncertain inclination can bias $M_\bullet$ by $30-50\%$ in elliptical galaxies \citep{Verolme,Geb03,Shapiro}.    
Moreover, isophotal evidence and simulations of galaxy mergers suggest that many BCGs are prolate or triaxial \citep{PSH91,RLP93,BKMaQ06}.  Although the orbit superposition method was originally developed for a triaxial potential \citep{Schild}, triaxial orbit models with the spatial and velocity resolutions necessary to measure $M_\bullet$ are a very recent development \citep{vdB08}.  In an early comparison of triaxial and axisymmetric models, \citet{vdB10} found that $M_\bullet$ was unchanged in M32, whereas triaxial models of NGC 3379 increased $M_\bullet$ by a factor of 2.

%
\section{Conclusions and Discussion}
\label{sec:disc}

We have reported the first stellar dynamical measurement of the central
black hole mass in a BCG beyond the Virgo Cluster.  Our results are based on
2-dimensional stellar kinematics in the central region of NGC 6086 (the
BCG in Abell 2162): the inner $0.9''$ (580 pc) from the IFS OSIRIS with
LGS-AO at Keck, the inner $4.9''$ from the IFS GMOS-N at Gemini North, and
long-slit data out to $30''$ from \citet{Carter}.  The individual datasets play complementary roles in constraining the gravitational potential of NGC 6086.  Used together, GMOS and OSIRIS data reduce degeneracy between the black hole mass and enclosed stellar mass near the center, decreasing the uncertainties of both quantities.  The long-slit data confirm the presence of dark matter and constrain the total enclosed mass, but they are insensitive to the precise form of the dark matter halo profile.  

We have used axisymmetric stellar orbit models
including a dark matter halo to determine $M_\bullet$ and the $R$-band
stellar mass-to-light ratio.  We have tested several dark matter halo profiles with our full set of kinematic measurements; in each case, the best-fit black hole is at least 4 times as massive as the best fit without dark matter.
Including dark matter in the models decreases the best-fit value of M$_\star$/L$_R$ by $20-40\%$.  For the most massive halo allowed
within the gravitational potential of the host cluster, we find $M_\bullet
= 3.6^{+1.7}_{-1.1} \times 10^9$ M$_\odot$ and M$_\star$/L$_R= 4.6^{+0.3}_{-0.7}$
M$_\odot$/L$_\odot$.  We obtain similar values of $M_\bullet$ and M$_\star$/L$_R$ when we exclude long-slit data, for models with and without dark matter.

\begin{figure}
      \epsfig{figure=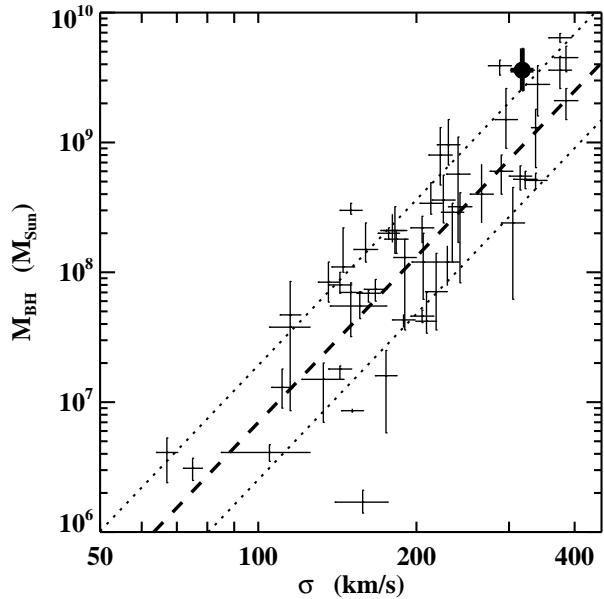,width=3.2in}
 \caption{The $M_\bullet - \sigma$ relationship.  The filled circle is our measurement of $M_\bullet = 3.6^{+1.7}_{-1.1} \times 10^9$ M$_\odot$ in NGC 6086, using the maximum-mass LOG dark matter halo ($v_c = 500$ km s$^{-1}$) and our full set of data from OSIRIS, GMOS, and \citet{Carter}.  Without dark matter, we measure $M_\bullet = 0.6^{+0.4}_{-0.4} \times 10^9$ M$_\odot$.  The remaining data points are the measurements compiled in \citet{Gultekin}, plus updated measurements
for M87 and M60 \citep{GT09,SG10}.  The thick dashed line is the fit  $\rm log \left(\it M_\bullet \right) = 8.12 + 4.24 \, \rm log \left( \sigma / 200 \, \rm km s^{-1} \right)$, and the dotted lines enclose a scatter of 0.44 dex, from \citet{Gultekin}}
\label{fig:Msigma}
\end{figure}

In Figure~\ref{fig:Msigma}, we add our measurement of $M_\bullet$ in NGC
6086 to the $M_\bullet - \sigma$ relationship of \citet{Gultekin}.  Our
plotted measurement corresponds to the most massive dark matter halo used
in our models.  We derive an effective velocity dispersion of $318 \pm 2$
km s$^{-1}$ within 1 effective radius, by weighting the kinematic measurements
from \citet{Carter} with respect to our measured surface brightness
profile.  The $M_\bullet - \sigma$ relation and intrinsic scatter from
\citet{Gultekin} yield a $\sigma$-predicted black hole mass of
$0.9^{+1.7}_{-0.6} \times 10^9$ M$_\odot$.
Our measurement of $M_\bullet = 3.6^{+1.7}_{-1.1} \times 10^9$ M$_\odot$
with a maximum-mass dark matter halo is marginally consistent with this
prediction, while our measurement of $M_\bullet = 2.6^{+1.3}_{-1.0} \times
10^9 M_\odot$ for an intermediate-mass halo is closer to the predicted
value.  
Our measurement of $M_\bullet = 0.6 \pm 0.4 \times 10^9$ M$_\odot$ without dark matter also agrees with the predicted value.  However, the unrealistic omission of dark matter at radii covered by long-slit data biases $M_\bullet$ toward low values.  
The $V$-band luminosity of NGC
6068 is $1.4 \times 10^{11}$ L$_{\odot,V}$, using $M_V = -23.11$ from
\citet{Lauer07}.  The $M_\bullet - L$ relation and intrinsic scatter from
\citet{Gultekin} yield a prediction of $M_\bullet = 1.3^{+1.9}_{-0.7}
\times 10^9$ M$_\odot$,
which is consistent with our measurements with and without dark matter.  

%

Of the existing sample of $\sim 30$ galaxies with $M_\bullet$ measured from
stellar dynamics, this study of NGC 6086 is only the third to consider dark
matter.  Like NGC 6086, measurements of $M_\bullet$ and M$_\star$/L in M87
depend strongly on the inclusion of a dark matter halo \citep{GT09}.
However, \citet{SG10} found that introducing dark matter to models of M60
produced minimal changes in $M_\bullet$ and M$_\star$/L.  Several factors
contribute to the greater importance of dark matter in models of NGC 6086
and M87.  First, the quality and spatial resolution of the kinematics are
insufficient to distinguish thoroughly the black hole from the central stellar mass profile.  Neither NGC 6086 nor M87 have good
spectra from \textit{HST}, and the potential advantage of AO IFS data for
NGC 6086 was compromised by low $S/N$ and coarse spatial binning.  In principle, very high-quality data could reveal unambiguously a compact mass at the center of the gravitational potential, permitting robust measurements of $M_\bullet$ in spite of biased M$_\star$/L$_R$ values.  Second,
shallow stellar mass profiles in NGC 6086 and M87 limit the range of radii
where stars dominate the gravitational potential, especially in the
presence of a relatively massive and concentrated dark matter halo.  In
contrast, models of M60 can exclude dark matter and still accurately
constrain M$_\star$/L by removing only the few outermost data points.  Third, shallow stellar light profiles affect kinematic measurements near the centers of NGC 6086 and M87: there is more contamination along the line
of sight from stars at larger radii, where dark matter is influential.
The latter two factors suggest that existing stellar dynamical measurements of
$M_\bullet$ are most likely biased in galaxies with large stellar cores.
New and revised measurements using models with dark matter could raise or steepen
the upper end of the black hole scaling relations, as stellar core size
increases with galaxy mass \citep{Lauer07}.    

Stellar population modeling can yield an independent measurement of the stellar mass-to-light ratio.  To compare existing stellar population studies to our dynamical results, we translate M$_\star$/L$_V$ and M$_\star$/L$_I$ to $R$ band by using the ($g-r$) and ($r-i$) colors of NGC 6086 from the Sloan Digital Sky Survey (SDSS).  Applying the filter translations of \citet{BR07}, we find $V-R = 0.64$ and $R-I = 0.68$; the resulting adjustments in solar units are M$_\star$/L$_R$ = 0.76 M$_\star$/L$_V$, and M$_\star$/L$_R$ = 1.33 M$_\star$/L$_I$.  
For the maximum-mass dark matter halo,
our dynamical measurement of M$_\star$/L$_R = 4.6^{+0.3}_{-0.7}$ M$_\odot$/L$_{\odot,R}$ 
agrees with population-based measurements from \citet{Capp06}, who find M$_\star$/L$_I \approx 3$ M$_\odot$/L$_{\odot,I}$ (M$_\star$/L$_R \approx 4$ M$_\odot$/L$_{\odot,R}$) for M87 and several other early-type galaxies observed with SAURON.  
However, a different range of stellar mass-to-light ratios is suggested by the study of \citet{vdLinden}, who model the stellar masses of 625 BCGs from SDSS.  The peak values of their mass and luminosity distribution functions yield a ratio of M$_\star$/L$_V = 3.1$ M$_\odot$/L$_{\odot,V}$ (M$_\star$/L$_R = 2.4$ M$_\odot$/L$_{\odot,R}$).  Additionally, \citet{GF10} have compiled various estimates of M$_\star$/L for a sample of $\sim 16,000$ early-type galaxies from SDSS, finding that M$_\star$/L increases with $\sigma$, and M$_\star$/L$_V \approx 2-3$ M$_\odot$/L$_{\odot,V}$ (M$_\star$/L$_R \approx 1.5-2.3$ M$_\odot$/L$_{\odot,R}$) for the highest-dispersion objects in their sample ($\sigma \sim 250$ km s$^{-1}$).  Both of these results fall significantly below our dynamical values of M$_\star$/L$_R$ for NGC 6086.  The discrepancy is most severe for orbit models without dark matter (M$_\star$/L$_R = 6.7^{+0.2}_{-0.9}$ M$_\odot$/L$_{\odot,R}$).  

To test whether any model of the gravitational potential for NGC 6086 can hold lower values of M$_\star$/L$_R$ and still fit our data, we have run a series of orbit models with M$_\star$/L$_R$ fixed at 2.5 M$_\odot$/L$_{\odot,R}$.  We sampled the LOG dark matter halo parameters $v_c$ and $r_c$ over a wide range of values and marginalized $\chi^2$ over trials with $M_\bullet = 3.5 \times 10^9$ M$_\odot$ and $M_\bullet = 7.0 \times 10^9$ M$_\odot$.  The resulting best-fit parameter values are $v_c =  420$ km $s^{-1}$ and $r_c$ = 2.0 kpc.  Fixing $v_c$ and $r_c$ at these values and finely sampling $M_\bullet$ yields $M_\bullet = 5.5^{+2.0}_{-0.9} \times 10^9$ M$_\odot$, including systematic errors, with $\chi^2_{min} = 1018.6$.  In comparison, $\chi^2_{min} = 1010.2$ for our maximum-mass LOG halo ($v_c = 500$ km s$^{-1}$; $r_c = 8.0$ kpc).  We conclude that the lower values of M$_\star$/L$_R$ motivated by stellar population modeling can produce a reasonable fit to our kinematics, given a more centrally concentrated dark matter halo.  Nonetheless, our original assumptions about the dark matter halo profile of NGC 6086 produce a better fit than the assumption of low M$_\star$/L$_R$.  Decreasing the enclosed stellar mass leads to a larger best-fit black hole mass, matching the trend from our other trials.

For each individual dark matter halo, the dominant systematic effect in our results is a discrepancy in the wings of LOSVDs from IFS versus long-slit data, illustrated by a jump in $h_4$ from negative to
positive values.  The magnitude of the resulting error is $22\%$ in $M_\bullet$ and
$9\%$ in M$_\star$/L$_R$.  Uncertainties in determining an optimal template star and
AO PSF yield smaller errors, totaling $\sim 9\%$ in $M_\bullet$ and $\sim
2\%$ in M$_\star$/L$_R$.  Additional, unmeasured systematic errors may
arise from our assumptions of axisymmetry and edge-on inclination.

Our investigation of NGC 6086 has established that stellar-dynamical
measurements of $M_\bullet$ in BCGs are possible with existing facilities;
this dramatically expands the sample volume of viable targets.  We plan to
follow this work with similar measurements from an ongoing survey of several
additional BCGs.  This survey is a critical step toward a statistically robust census of
black holes in the Universe's most massive galaxies, and will eventually
shed new light on the histories of galaxies and black holes at the hearts
of galaxy clusters.\\

We are grateful for insightful discussions with many colleagues during this investigation: particularly, Genevieve Graves, Kayhan G\"{u}ltekin, Anne Medling, Kristen Shapiro, Remco van den Bosch, and Jong-Hak Woo.  We also thank Sandra Faber and Scott Tremaine for their contributions to our OSIRIS observing proposal.  Support astronomers Randy Campbell, Al Conrad, and Jim Lyke played crucial roles in conducting observations with OSIRIS.  We thank David Carter and George Hau for providing long-slit data of NGC 6086.  We thank the anonymous referee for many helpful suggestions.

This work was supported by NSF AST-1009663.  Work by N.J.M. and J.R.G. was supported in part by the NSF Center for Adaptive Optics, managed by the University of California at Santa Cruz under cooperative agreement AST 98-76783.  Support for C.-P.M. is provided in part by the Miller Institute for Basic Research in Science, University of California, Berkeley.  KG acknowledges support from NSF-0908639.  This investigation would not have been possible without the facilities at the Texas Advanced Computing Center at The University of Texas at Austin, which has allowed access to over 5000 node computers were we ran all of the models.  The W. M. Keck Observatory is operated as a scientific partnership among the California Institute of Technology, the University of California, and the National Aeronautics and Space Administration. The Observatory was made possible by the generous financial support of the W. M. Keck Foundation.  The Gemini Observatory is operated by the 
Association of Universities for Research in Astronomy, Inc., under a cooperative agreement 
with the NSF on behalf of the Gemini partnership: the National Science Foundation (United 
States), the Science and Technology Facilities Council (United Kingdom), the 
National Research Council (Canada), CONICYT (Chile), the Australian Research Council (Australia), 
Minist\'{e}rio da Ci\^{e}ncia e Tecnologia (Brazil) 
and Ministerio de Ciencia, Tecnolog\'{i}a e Innovaci\'{o}n Productiva (Argentina).  GMOS data of NGC 6086 were obtained under Program GN-2003A-Q-11.

In addition, the authors wish to recognize and acknowledge the very significant cultural role and reverence that the summit of Mauna Kea has always had within the indigenous Hawaiian community.  We are most fortunate to have the opportunity to conduct observations from this mountain.







\appendix
\makeatletter
\def\@seccntformat#1{Appendix\ \csname the#1\endcsname\quad}
\makeatother

%
\section{A: Uncertainties in the PSF and Stellar Templates for OSIRIS}
\label{app:psf}

We ran additional series of orbit models to assess the effects of stellar template mismatch and PSF uncertainty on the best-fit values of $M_\bullet$ and M$_\star$/L$_R$.  We expect these systematic effects to be largest for OSIRIS data.  Lower signal-to-noise spectra have larger uncertainties in template matching, and the crowded series of atomic and molecular features in $H$-band is more sensitive to template choice than the CaII features at 0.85 $\mu$m.  The structure of the LGS-AO PSF is sensitive to atmospheric turbulence, laser power, the density and thickness of the ionospheric sodium layer, performance of the wavefront sensor and deformable mirror, and the brightness and position of the tip/tilt star.  Because several of these factors change over time, delays between observations of an extended science target and a point source can induce errors the estimated PSF.  Uncertainties in seeing-limited PSFs arise predominantly from changes in atmospheric turbulence.
In order to highlight the effect of template and PSF errors for OSIRIS, we excluded the more stable GMOS data from our trials below.  Each trial included our maximum mass dark matter halo.  We summarize the trials in Table~\ref{tab:Otest}.

%
\begin{table*}[htbp]
\begin{center}
\caption{PSF and Template Star Trials}
\label{tab:Otest}
\begin{tabular}[b]{ccccccccc}  
Data & PSF & Template & $v_c$ & $r_c$ & $M_\bullet$  & M$_\star$/L$_R$ & $\chi^2_{\rm min}$ & $N_{dof}$ \\
&&& (km $s^{-1}$) & (kpc) & ($10^9 \; \rm M_\odot$) & (M$_\odot$/L$_\odot$) & & \\
\\
(1) & (2) & (3) & (4) & (5) & (6) & (7) & (8) & (9)\\
\hline 
\\
OSIRIS + \citet{Carter} & A  & M4III & 500 & 8.0 & $1.8^{+1.3}_{-1.0}$ & $4.5^{+0.7}_{-0.6}$ & 419.0 & 570\\
\\
OSIRIS + \citet{Carter} & B & M4III & 500 & 8.0 & $1.9^{+1.4}_{-1.1}$ & $4.6^{+0.7}_{-0.7}$ & 412.7 & 570\\
\\
OSIRIS + \citet{Carter} & B & M0III & 500 & 8.0 & $2.0^{+1.3}_{-1.0}$ & $4.5^{+0.7}_{-0.7}$ & 415.6 & 570\\
\\
OSIRIS + \citet{Carter} & C & M4III & 500 & 8.0 & $2.1^{+1.4}_{-1.1}$ & $4.5^{+0.7}_{-0.6}$ & 414.6 & 570\\
\\
\hline
\end{tabular}
\end{center}
\begin{small}
\textbf{Notes:}  Column 1: Data sets included in trial.  Column 2: estimated PSF for OSIRIS data.  A: original PSF measured from tip/tilt star with OSIRIS, folded over major and minor axes of NGC 6086 (Figure~\ref{fig:spatial}, top right).  B: tapered and folded PSF (Figure~\ref{fig:spatial}, bottom left).  C: core-halo PSF, with 25\% Strehl ratio (Figure~\ref{fig:spatial}, bottom right).  Column 3: spectral type of template star for OSIRIS data.  We used spectra from HD 110964 (M4III) and HD 108629 (M0III).  Column 4: circular velocity of LOG dark matter halo (Eq.~\ref{eq:halo}).  Column 5: core radius of dark matter halo (Eq.~\ref{eq:halo}).  Column 6: best-fit black hole mass.  Column 7: best-fit $R$-band stellar mass-to-light ratio.  Column 8: minimum $\chi^2$ value for all models.  Column 9: degrees of freedom in model fits to LOSVDs.  Computed values include a smoothing factor of 1 degree of freedom per 2 velocity bins for non-parametric LOSVDs from OSIRIS.  Quoted errors in $M_\bullet$ and M$_\star$/L$_R$ correspond to 68\% confidence intervals.  All trials in Section~\ref{sec:results} used PSF B and template star HD 110964 (M4III).
\end{small}
\end{table*}

\begin{figure}[htbp]
 \centering
 \epsfig{figure=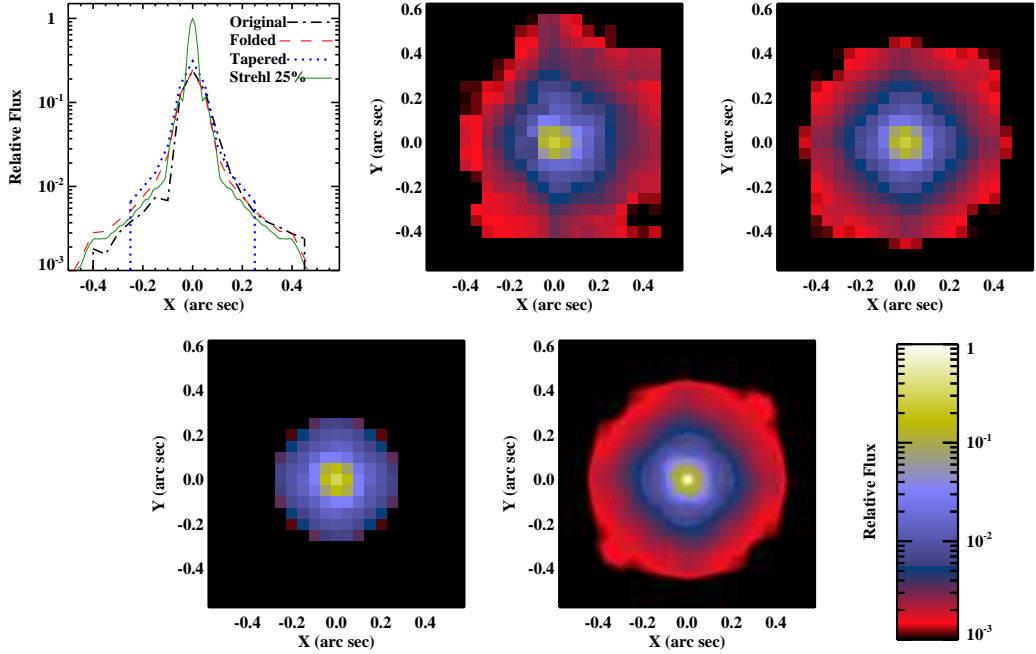,width=5.5in}
 \caption{PSFs estimated for OSIRIS observations of NGC 6086.  \textit{Top left:} horizontal slice through the center of each PSF.  \textit{Top middle:} Original estimate from the LGS-AO tip/tilt star.  \textit{Top right:} PSF A, folded for axisymmetry.  \textit{Bottom left:} PSF B, folded and tapered.  \textit{Bottom right:}  PSF C, interpolated for a diffraction-limited core and 25\% Strehl ratio.  The major and minor axes of NGC 6086 are oriented $45^\circ$ from the plotted $X$- and $Y$-axes.  Relative flux is defined with respect to the peak value of PSF C.  All of our orbit models used PSF B with OSIRIS data, with the exception of trials to test model dependence on PSF structure.}
\label{fig:spatial}
\end{figure}

In Figure~\ref{fig:spatial}, we display various estimates of the average PSF for OSIRIS observations of NGC 6086.  Our initial estimate, displayed at the top middle, was constructed from a one-time sequence of exposures of the LGS-AO tip/tilt star.  We have characterized the two-component structure of this PSF by fitting a narrow Gaussian profile plus a broad Moffat profile.  The resulting FHWM values are   $0.10''$ and $0.42''$.  The narrow component contains $44\%$ of the total flux.  However, we cannot directly compute a Strehl ratio from this percentage, because the $0.05''$ pixels in our collapsed OSIRIS images undersample the $H$-band diffraction limit ($\lambda/D = 0.033''$).  Our trials use PSFs derived from the original tip/tilt images, rather than the Gaussian plus Moffat model.  The PSFs used in our trials (A, B, and C) were all symmetrized with respect to the major- and minor-axis position angles of NGC 6086, in order to match the spatial folding of kinematic data for axisymmetric models.  

PSF A (Figure~\ref{fig:spatial}, top right) symmetrizes the estimate from the tip/tilt star, but contains no further changes.  This estimate has two primary limitations.  It only captures the PSF at a single moment in time, and it does not account for the $42''$ separation between the tip/tilt star and the center of NGC 6086.  A complementary method for estimating the PSF is to compare the collapsed OSIRIS mosaic of the galaxy center to an image from \textit{HST}/WFPC2.  We convolved this mosaic with an appropriate WFPC2 PSF and convolved the WFPC2 image with a trial PSF, so that both images would have the same total smoothing kernel.  Using our original PSF estimate from the tip/tilt star, we found that the brightness profile of the convolved WFPC2 image was shallower than the collapsed OSIRIS mosaic.  Therefore, we tested a series of PSFs that multiplied this original estimate by tapering functions with various inner and outer radii.  We found the lowest root-mean-squared residual for PSF B, depicted in Figure~\ref{fig:spatial} (bottom left), which suppresses power at radii beyond $0.3''$.  In order to probe the effect of a Nyquist-sampled, diffraction-limited core, we constructed a third estimate, PSF C (Figure~\ref{fig:spatial}, bottom right).  We interpolated the original tip/tilt-based PSF from $0.05''$/pixel to $0.01''$/pixel, and replaced the central $0.1'' \times 0.1''$ with a 2-dimensional Airy profile ($\lambda/D = 0.033''$).  The amplitude of the core was scaled relative to the outer profile to yield a Strehl ratio of 25\%.  

PSFs A, B, and C incorporate a variety of methods for PSF estimation, and the range of variation in their structures is comparable to the uncertainty in any particular method.  We find a standard deviation of 8\% in $M_\bullet$ and 1\% in M$_\star$/L$_R$ among trials using each of these three PSFs.  Similarly, \citet{Nowak08} found little variation in $M_\bullet$ and M$_\star$/L$_{K_s}$ from different estimates of the PSF for AO data of Fornax A.  For all of our primary trials to measure $M_\bullet$ and M$_\star$/L$_R$ in NGC 6086 (Section~\ref{sec:results}), we matched OSIRIS data to PSF B.  For data from GMOS and \citet{Carter}, we use Gaussian PSFs.  The GMOS PSF has a FWHM of $0.4''$, based on acquisition camera images.  \citet{Carter} report a FWHM of $1.3''$.  Uncertainties in these seeing-limited PSFs should have smaller effects on measurements of $M_\bullet$, due to poorer spatial resolution.

\begin{figure}[htbp]
 \centering
 \epsfig{figure=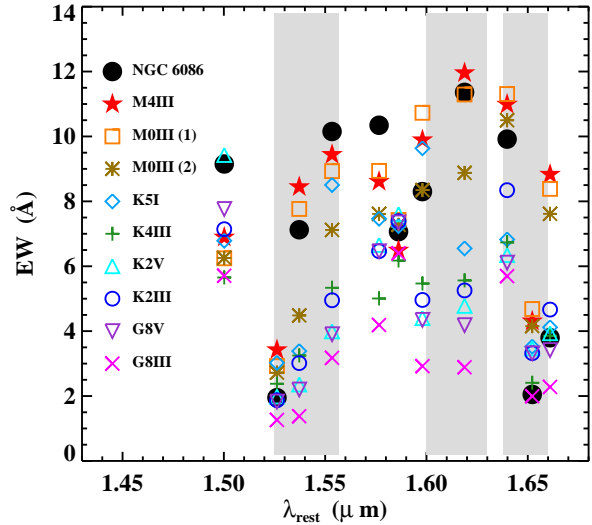,width=3.5in}
 \caption{Equivalent widths of NGC 6086 and template stars for different $H$-band spectral features.  The shaded areas mark the spectral range used for kinematic extraction from OSIRIS spectra.  Template stars are: HD 110964 (M4III); HD 108629 (M0III 1); HD 63348 (M0III 2); HD 44537 (K5I); 15 Lib (K4III); HD 99492 (K2V); HD 94386 (K2III); 55 Cnc (G8V); HD 89638 (G8III).}
\label{fig:EW}
\end{figure}

Our stellar template library for OSIRIS contains stars with spectral types from G8 through M4.  In principle, we can determine a best-fit weighted combination of library templates for each spectrum of NGC 6086, while simultaneously fitting for the LOSVD.  However, noise in our OSIRIS spectra renders this method unstable: when we include a large number of templates, the weights vary dramatically between adjacent spatial regions.  To narrow our sample, we compared the equivalent widths (EWs) of $H$-band absorption features in NGC 6086 to each library template, as illustrated in Figure~\ref{fig:EW}.  EWs for NGC 6086 were measured from a high-$S/N$ spectrum covering a large portion of the OSIRIS field-of-view.  Our kinematic fitting is dominated by high-EW features at 1.54 $\mu$m, 1.55 $\mu$m, 1.62 $\mu$m, and 1.64 $\mu$m.  For these features, the three M-giant stars provide much closer EW matches than other stars in our library.  Trial kinematic fits using only M-giant stars weighted HD 110964 (M4III) by nearly 100\%; for simplicity, we adopted this star as our only template for the final kinematic extraction from OSIRIS spectra.

To test template mismatch, we extracted a second set of LOSVDs from OSIRIS spectra, this time using HD 108629 (M0III) as the template star.  The difference between spectral types M0III and M4III is a reasonable estimate for template mismatch in our kinematic fitting, as indicated by our equivalent width measurements.  In particular, our fits exclude the Mg/Fe absorption feature near 1.50 $\mu$m rest, which is a better match to our K-dwarf template.  Orbit model trials with M4III-based LOSVDs versus M0III-based LOSVDs differ by 5\% in $M_\bullet$ and 2\% in M$_\star$/L$_R$.

Even though HD 110964 compares favorably to other template stars, its average equivalent width over the spectral features used to extract LOSVDs is $\approx 13\%$ higher than the average equivalent width of NGC 6086.  To improve the EW agreement, we artificially decreased the line strengths of the template spectrum, by a constant factor $f_{EW} = 0.83$.  We did not allow $f_{EW}$ to vary over different spatial regions in NGC 6086.
Optimizing the value of $f_{EW}$ produces smoother LOSVDs, which are also more stable to small changes in spectral smoothing.  However, we have found a positive correlation between $f_{EW}$ and the value of $h_4$ derived from the resulting fit.  In non-parametric terms, the wings of the best-fit LOSVD become truncated as absorption features in the template star are artificially made shallower.  This effect may contribute partially to the differences between our computed values of $h_4$ and those from \citet{Carter}.  

\clearpage

%
\section{B: Non-Parametric LOSVDs from OSIRIS and GMOS}
\label{app:losvd}

\begin{figure}[htbp]
 \centering
 \epsfig{figure=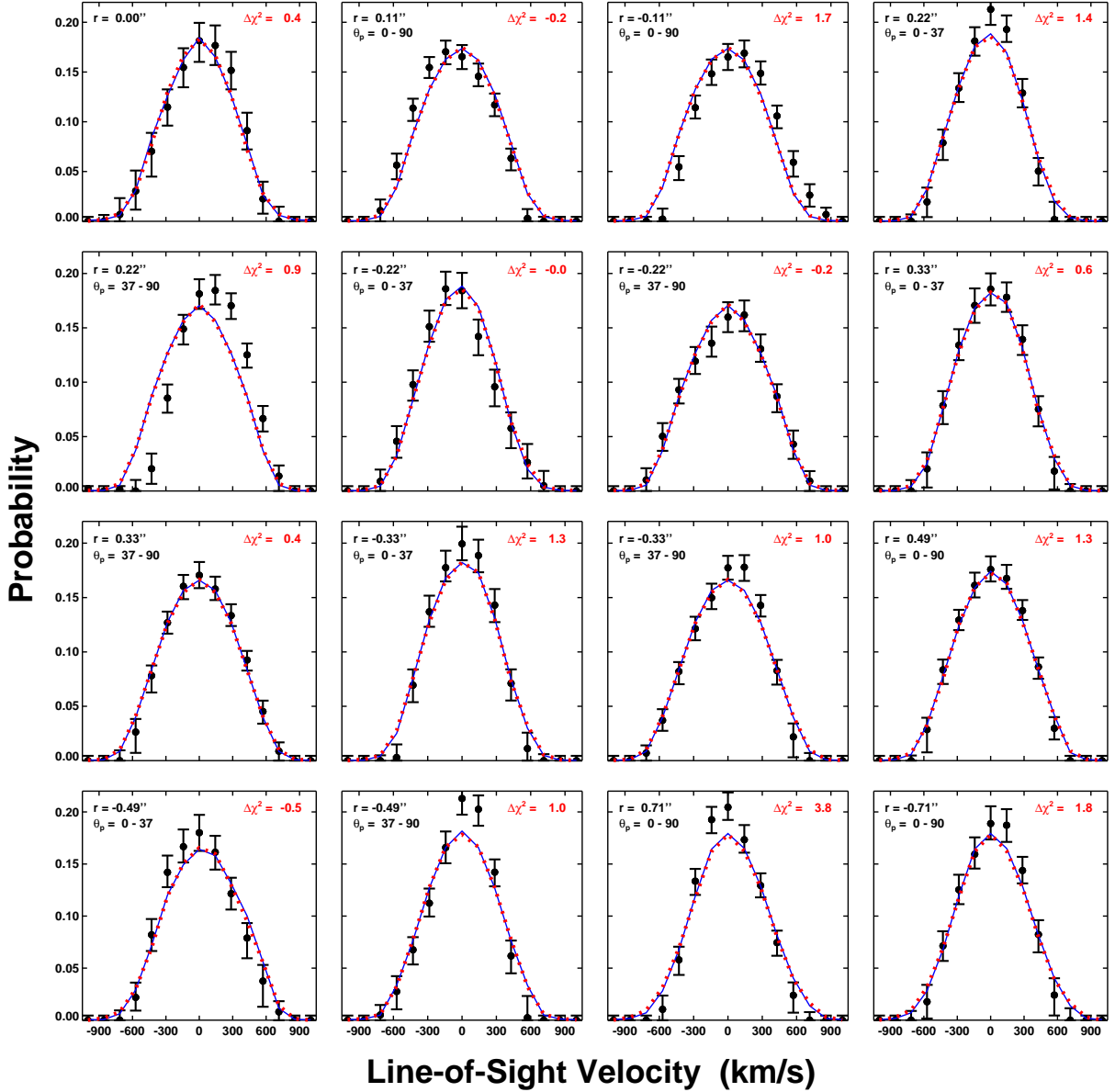,width=6.5in}
 \caption{LOSVDs from OSIRIS spectra.  Solid blue lines are corresponding LOSVDs generated by the best-fitting orbit model with the maximum-mass LOG dark matter halo (M$_\star$/L$_R$ = 4.7 M$_\odot$/L$_\odot$, $M_\bullet = 3.5 \times 10^9$ M$_\odot$, $v_c = 500$ kpc, $r_c = 8.0$ kpc).  Dotted red lines are generated by the best-fitting orbit model with no dark halo (M$_\star$/L$_R$ = 6.8 M$_\odot$/L$_\odot$, $M_\bullet = 3.2 \times 10^8$ M$_\odot$).  In each sub-plot, $r$ is the distance from the center of NGC 6086, and $\theta_p$ is the range of angles with respect to the major axis.  Negative values of $r$ indicate spectra from the south side of NGC 6086.  For each LOSVD, $\Delta\chi^2$ is the difference in the $\chi^2$ statistic for the two models: $\Delta\chi^2 > 0$ indicates that the model including dark matter (solid blue line) is a better fit.}
\label{fig:Olosvd}
\end{figure}

\begin{figure}[htbp]
 \centering
  \epsfig{figure=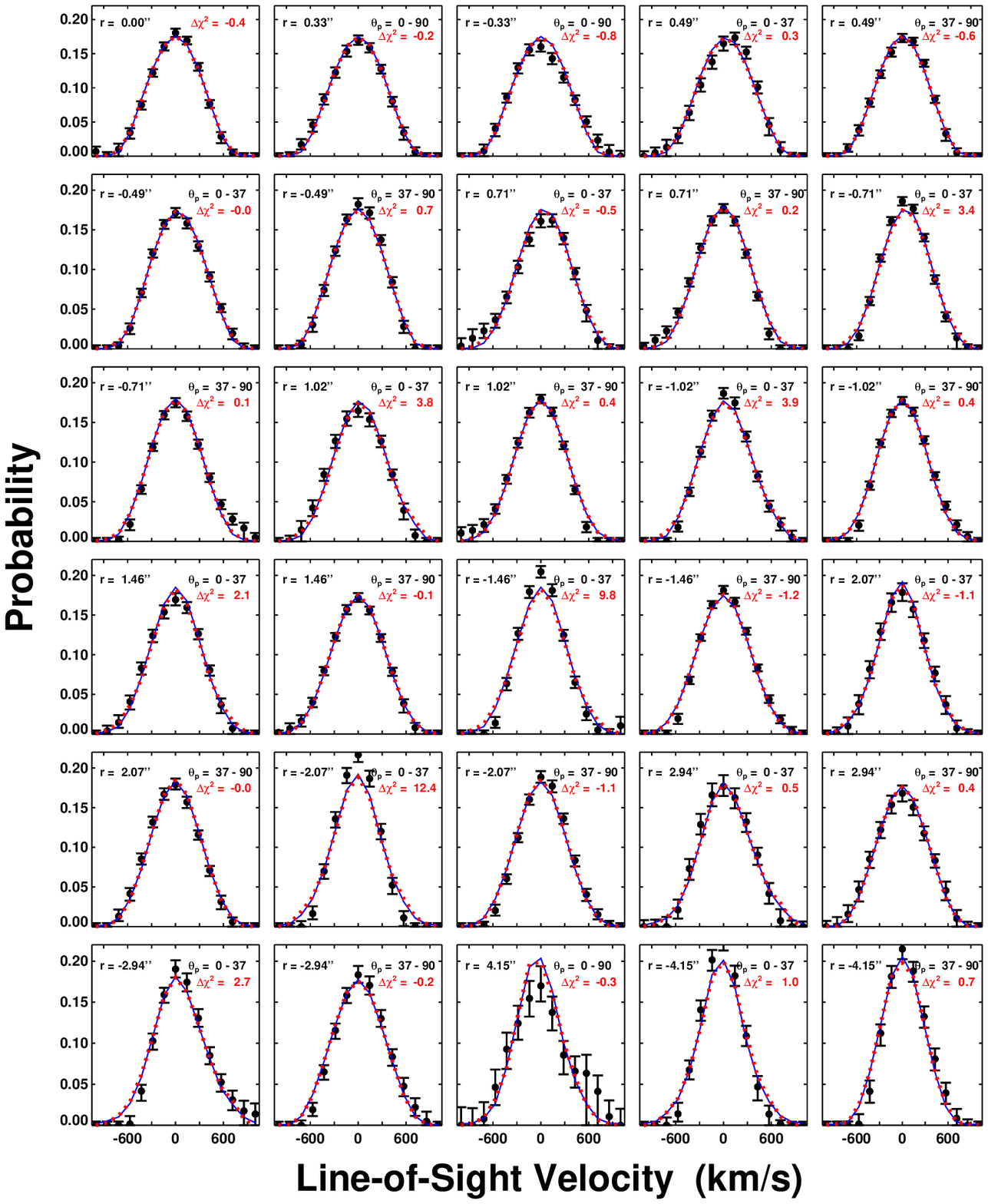,width=6.5in}
 \caption{LOSVDs from GMOS spectra.  Solid blue lines are corresponding LOSVDs generated by the best-fitting orbit model with the maximum-mass LOG dark matter halo (M$_\star$/L$_R$ = 4.7 M$_\odot$/L$_\odot$, $M_\bullet = 3.5 \times 10^9$ M$_\odot$, $v_c = 500$ kpc, $r_c = 8.0$ kpc).  Dotted red lines are generated by the best-fitting orbit model with no dark halo (M$_\star$/L$_R$ = 6.8 M$_\odot$/L$_\odot$, $M_\bullet = 3.2 \times 10^8$ M$_\odot$).  In each sub-plot, $r$ is the distance from the center of NGC 6086, and $\theta_p$ is the range of angles with respect to the major axis.  Negative values of $r$ indicate spectra from the south side of NGC 6086.  For each LOSVD, $\Delta\chi^2$ is the difference in the $\chi^2$ statistic for the two models: $\Delta\chi^2 > 0$ indicates that the model including dark matter (solid blue line) is a better fit.}
\label{fig:Glosvd}
\end{figure}

\begin{figure}[htbp]
 \centering
  \epsfig{figure=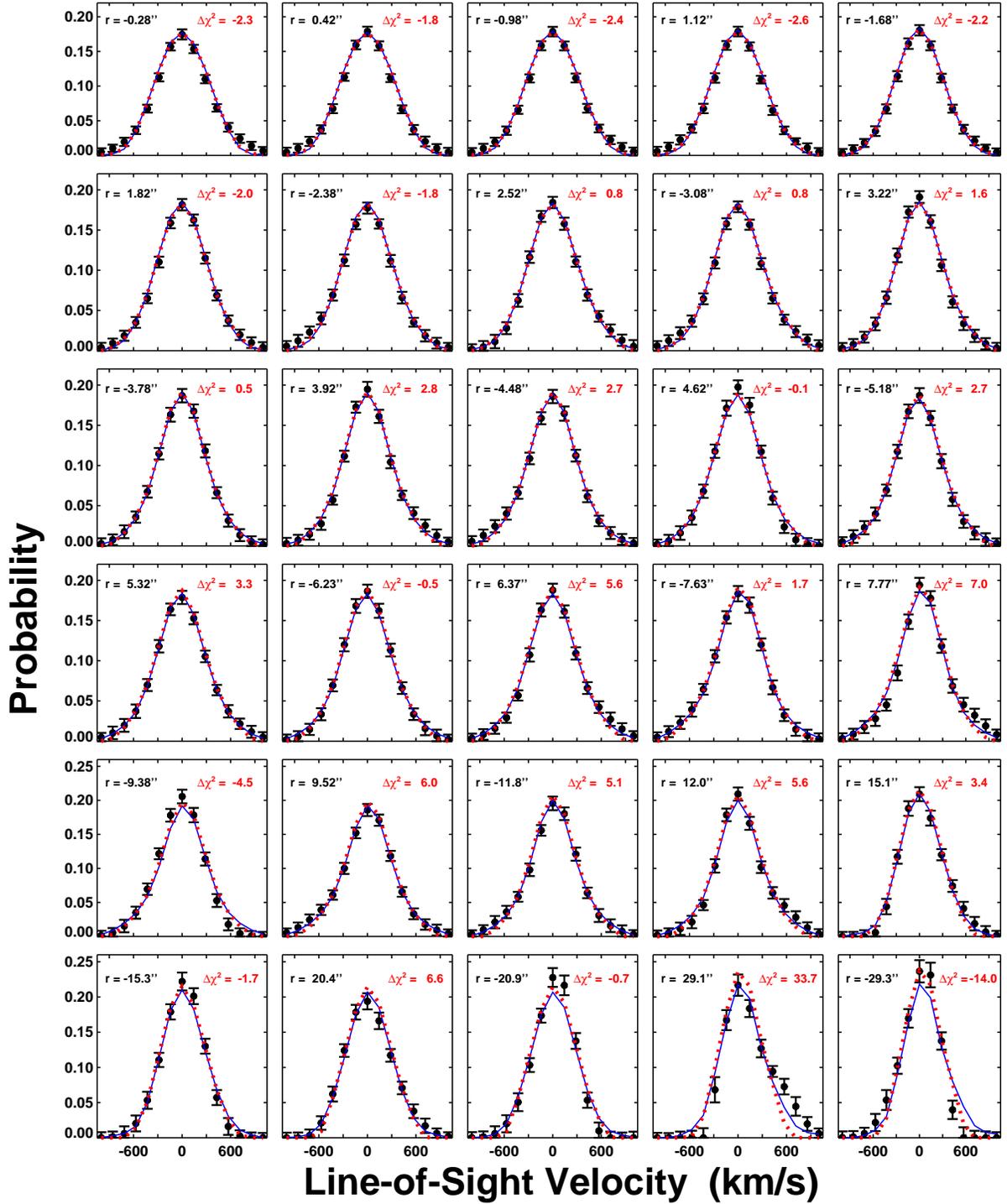,width=6.5in}
 \caption{LOSVDs generated from the kinematic moments measured by \citet{Carter}.  Solid blue lines are corresponding LOSVDs generated by the best-fitting orbit model with the maximum-mass LOG dark matter halo (M$_\star$/L$_R$ = 4.7 M$_\odot$/L$_\odot$, $M_\bullet = 3.5 \times 10^9$ M$_\odot$, $v_c = 500$ kpc, $r_c = 8.0$ kpc).  Dotted red lines are generated by the best-fitting orbit model with no dark halo (M$_\star$/L$_R$ = 6.8 M$_\odot$/L$_\odot$, $M_\bullet = 3.2 \times 10^8$ M$_\odot$).  In each sub-plot, $r$ is the distance from the center of NGC 6086, along the major axis.  Negative values of $r$ indicate spectra from the south side of NGC 6086.  For each LOSVD, $\Delta\chi^2$ is the difference in the $\chi^2$ statistic for the two models: $\Delta\chi^2 > 0$ indicates that the model including dark matter (solid blue line) is a better fit.}
\label{fig:Ilosvd}
\end{figure}

We present our final extracted LOSVDs from OSIRIS (Figure~\ref{fig:Olosvd}) and GMOS (Figure~\ref{fig:Glosvd}), and LOSVDs derived from the measurements of \citet{Carter} (Figure~\ref{fig:Ilosvd}).  The LOSVDs from OSIRIS use template star HD 110964 (M4III).
We compare each LOSVD to the best-fitting orbit model with our maximum-mass LOG halo, and the best-fitting orbit model with no dark matter.  Including dark matter gives a better fit to the total set of LOSVDs, with a cumulative difference $\Delta\chi^2 = 104.2$.  Similarities between the LOSVDs generated by each model occur in part because $\sim 30,000$ orbits give the models a high degree of flexibility to optimally fit the data.  The modelsÕ inability to perfectly match the data arises in part from constraints such as axisymmetry and uniformity in M$_\star$/L$_R$.   

%

%

%

\end{document}